\documentclass[10pt,aps,prd,twocolumn,superscriptaddress]{revtex4-2}

\usepackage{amsmath}
\usepackage{amssymb}
\usepackage{graphicx}% Include Figure~files
\usepackage{dcolumn}% Align table columns on decimal point
\usepackage{bm}% bold math
\usepackage{natbib} % for \citep and enhanced citation features
\usepackage{anyfontsize}
\usepackage{xcolor} 
\usepackage{ulem}
\usepackage{url}
\usepackage{booktabs,multirow,tabularx}% http://ctan.org/pkg/{booktabs,multirow,tabularx}
\usepackage{aas_macros}

\newcommand{\newtext}[1]{\textcolor{black}{#1}}

\begin{document}

\preprint{APS/123-QED}

%\title{Weak lensing statistic covariance replication with flow based models}% Force line breaks with \\
%\thanks{Correspondence email}%
\title{Replicating weak-lensing summary-statistic covariances with normalizing flows}

\author{Joaquin Armijo}
\email{joaquin.armijo@usp.br}
\affiliation{Center for Data-Driven Discovery, Kavli IPMU (WPI), UTIAS, The University of Tokyo, Kashiwa, Chiba 277-8583, Japan}
\affiliation{Department of Mathematical Physics, Institute of Physics, University of São Paulo, R. do Matão 1371, 05508-090, São Paulo, SP, Brazil}

\author{Leander Thiele}
\affiliation{Center for Data-Driven Discovery, Kavli IPMU (WPI), UTIAS, The University of Tokyo, Kashiwa, Chiba 277-8583, Japan}

\author{Jia Liu}
\affiliation{Center for Data-Driven Discovery, Kavli IPMU (WPI), UTIAS, The University of Tokyo, Kashiwa, Chiba 277-8583, Japan}

\begin{abstract}
We explore the ability of normalizing flow (NF) generative models to reproduce weak-lensing summary statistics when trained on a set of cosmological simulations. Our analysis focuses on how accurately NF models recover the mean, standard deviation, and covariance of key statistics derived from convergence ($\kappa$) maps: The angular power spectrum $C_{\ell}$, probability density function, and Minkowski functionals of weak lensing convergence $\kappa$-maps. We test two scenarios for training: (1) on the data vectors and (2) on the full $\kappa$-maps. In both cases, the NF models reproduce the mean and variance of the target statistics within percent-level accuracy. However, the accuracy of the off-diagonal elements of the covariance matrix is underestimated by up to $\sim25\%$. We study several mitigation strategies and find that data augmentation and training with noisy fields help improve covariance recovery to \newtext{$\mathcal{O}(5\%)$ on power spectrum statistics}. Our study demonstrates that while the means and variances of weak lensing statistics can be well modeled by NF, covariances can be significantly underestimated if mitigation strategies are not applied. \newtext{We present this test as a rigorous diagnostic of generative-model fidelity.}

\end{abstract}

\maketitle

%\tableofcontents

\section{Introduction} \label{sec:introduction}

\noindent Weak gravitational lensing is a powerful observational tool that probes the large-scale distribution of matter in the universe. It arises from the subtle distortion of the shapes of distant galaxies due to the gravitational influence of intervening mass along the line of sight~\cite{Bartelmann2001, Kilbinger2015,mandelbaum2018weak}. By statistically analyzing these distorted shapes across a large sample of galaxies, the distribution of matter can be inferred and studied over cosmic history. This makes weak lensing a key probe for inferring cosmological parameters, such as the matter energy-density $\Omega_{\rm m}$, the amplitude of matter fluctuations $\sigma_8$, and the properties of dark energy~\cite{Huterer2010,Huterer2018}, thereby offering critical insights into the fundamental nature of the universe.

Weak lensing fields are designed through observations of large extragalactic surveys, such as the Dark Energy Survey (DES)~\cite{DES2022}, the Kilo-degree survey \cite{Heymans2021}, the Hyper Suprime-Cam (HSC)~\cite{Hikage2019}, and upcoming ones, such as Euclid~\cite{Laureijs2011} and Vera C. Rubin Observatory Legacy Survey of Space and Time (LSST)~\cite{LSST2009} missions. These experiments have observed millions of distant galaxies across vast areas of the sky, enabling the creation of lensing catalogs by precisely measuring the cosmic shear of individual galaxies. From these shape measurements, the projected mass density can be reconstructed~\cite{KaiserSquires1993}, allowing researchers to produce detailed maps of the underlying matter distribution.

Cosmological simulations are essential for modeling the late-time matter field and generating mock universes with gravitational lensing effects~\cite{Takahashi2017}. These simulations help to test mass-mapping techniques~\cite{ValeWhite2003,Pires2009,Jeffrey2021}, model observational systematics \citep{Massey2013}, interpreting cosmic shear data \cite{harnois2018slics}, and including the effect of baryons~\cite{Troster2019}, ultimately improving our ability to constrain cosmological models. However, producing large ensembles that simulate the large volume and high-resolution of the data remains computationally expensive and has become a major bottleneck for weak-lensing analyses. Machine-learning–based generative models offer a promising path forward. Methods such as VAEs, GANs, NFs, and diffusion models can learn the non-Gaussian structure of weak-lensing convergence maps directly from simulations~\cite{Rodriguez2018,Ravanbakhsh2017,Mustafa2019,Fluri2019,Dai_Seljak2023,Whitney2024,Whitney2024b,Boruah2025}. Once trained, these models can generate high-fidelity synthetic $\kappa$-maps at negligible computational cost, reproducing both standard and higher-order statistics~\cite{Gupta2018}. This enables extensive mock catalog production, exploration of rare events, and scalable uncertainty quantification. When conditioned on cosmological parameters, they additionally serve as fast emulators for cosmological inference~\cite{Perraudin2020,Fluri2019,Remy2020}, providing a data-driven complement to traditional simulation pipelines.

%\jl{moved the next 2 paragraphs from section 2 to here, maybe missed placed before..}

In this work, we study flow-based generative models as a flexible and interpretable framework for modeling weak-lensing convergence maps generated by cosmological simulations. Normalizing flows (NFs) learn an invertible mapping between a multivariate Gaussian latent space and the target data distribution, providing an explicit and tractable likelihood and avoiding mode collapse, in contrast to adversarial approaches~\citep{Dai2022, Dai_Seljak2023}. When trained on suites of simulations, NFs approximate the probability distribution of convergence fields and enable the generation of large ensembles of high-fidelity synthetic maps with the same dimensionality as the original data. Given the variable performance of neural networks in cosmological applications, their validity must be assessed by how accurately they reproduce key summary statistics and utilize the available training information. We apply NFs to two-dimensional convergence ($\kappa$) fields, which are line-of-sight projections of the matter density, and evaluate statistical fidelity using complementary probes: the angular power spectrum $C_{\ell}^{\kappa\kappa}$, which captures Gaussian information, the one-point $\kappa$ probability distribution function, which probes higher-order moments, and Minkowski functionals, which characterize the topology of the field. \newtext{We emphasize that our use of the covariance matrix is diagnostic in nature. The aim is not to propose NFs as a replacement for simulation-based covariance estimation: when a thousand simulation suite is already available, the empirical covariance is by construction accurate for a required summary statistics. Rather, the covariance is a sensitive probe of whether the NF network has correctly learned the joint distribution of summary statistic bins (or map-level pixel correlations). Mean and variance can be matched by simple generative models, whereas the off-diagonal covariance recovery is a more demanding test. Establishing whether NF-generated maps reproduce this structure is a prerequisite for their use as fast emulators, mock generators for rare-event studies, or components of simulation-based inference pipelines.} 

%This is especially valuable for estimating covariance matrices, which typically require thousands of expensive simulations to compute accurately~\cite{Kodwani2023,HallTaylor2019,Oconnell2019,Petri2016,SellentinHeavens2017}. 
%By augmenting the dataset with generated samples that retain the correct statistical properties, these models enable more efficient and stable covariance estimation. This, in turn, improves the reliability of cosmological parameter inference while significantly reducing the computational cost associated with traditional simulation-based approaches. We intent to test how valid NF techniques are to replicate trustful information, specially the covariance, and to provide insight on how the accuracy of the replicated information scales with dimensionality. \newtext{In this paper we will try to illustrate how to improve the accuracy of the replicated information by conserving (roughly) the size of the network, which will give us an idea of how big the network should be for a determined setup (for example to replicate maps with sub-arcmin resolution with large size)... Need to reword.}.

This paper is organized as follows: We describe weak lensing, including convergence fields and their summary statistics in section \ref{sec:WL}. The simulations used to train the neural network are described in Section \ref{sec:simulations}. A description of normalizing flow networks is written in Section \ref{sec:NF}. We show and discuss the results of the replication of summary statistics in \ref{sec:results}. Finally, conclusions are drawn in Section \ref{sec:conclusions}.

\section{Weak Lensing} \label{sec:WL}

 The deflection of extragalactic background source light by a gravitational potential leads to coherent distortions in the observed galaxy shape, a phenomenon known as cosmic shear. The weak lensing effect is described by the Jacobian matrix $\mathcal{A}$, which relates the unlensed position of a light ray to its observed (lensed) position:

\begin{equation}
\mathcal{A} =
\begin{pmatrix}
1-\kappa-\gamma_1 & -\gamma_2 \\
-\gamma_2         & 1-\kappa+\gamma_1
\end{pmatrix}
\end{equation}

where $\kappa$ denotes the convergence, which captures the isotropic magnification of images, and $\gamma = \gamma_1 + i\gamma_2$ is the complex \textit{shear}, responsible for the anisotropic distortion of galaxy shapes. In the weak-lensing regime, where $|\kappa|, |\gamma| \ll 1$, the lensing transformation is well described by a linear approximation. Since the lensing-induced shear of any individual galaxy is much smaller than its intrinsic shape noise, the signal must be statistically recovered by averaging over large ensembles of galaxies.

The convergence $\kappa$ is of particular interest because it directly relates to the projected matter density along the line of sight. It is defined as a weighted integral of the three-dimensional matter overdensity field $\delta$:

\begin{equation}
\kappa(\boldsymbol{\theta}) = \int_0^{\chi_s} \mathrm{d}\chi W(\chi)  \delta\left( \chi, \boldsymbol{\theta}\right),
\end{equation}

where $\chi$ is the comoving distance, $\chi_s$ is the distance to the source galaxy, and $W(\chi)$ is the lensing efficiency kernel, given by:

\begin{equation}
W(\chi) = \frac{3H_0^2\Omega_m}{2c^2} \frac{\chi}{a(\chi)} \int_\chi^{\chi_s} \mathrm{d}\chi' n(\chi') \frac{\chi' - \chi}{\chi'},
\end{equation}

with $n(\chi')$ being the source galaxy distribution, $c$ the speed of light, $a(\chi)$ the scale factor, $H_0$ the Hubble constant, and $\Omega_m$ the matter density parameter.

Mapping the convergence field provides a two-dimensional view of the projected matter distribution, which includes both dark and baryonic matter. These convergence maps are key for studying the growth of cosmic structure and for inferring cosmological parameters.

Recent and ongoing weak lensing surveys have significantly advanced our ability to create high-fidelity convergence maps and perform statistical inference. The Dark Energy Survey (DES) \cite{DES2021cosmicshear}, the Kilo-Degree Survey (KiDS) \cite{asgari2021kids}, and the Hyper Suprime-Cam (HSC) survey \cite{hikage2019hsc} have provided precise measurements of cosmic structure.

\subsection{Summary statistics}

Extracting cosmological information from weak-lensing convergence maps relies on summary statistics that quantify the spatial structure of the projected matter density field. Although $\kappa$ is a line-of-sight projection of the underlying matter distribution, it becomes significantly non-Gaussian on small angular scales as a result of nonlinear gravitational evolution. Consequently, a broad range of summary statistics is employed to capture the information content of weak-lensing fields, spanning traditional two-point statistics and higher-order statistics providing significant non-Gaussian information.

\subsubsection*{Convergence Angular Power Spectrum}

The angular power spectrum of the convergence field, $C_\ell^{\kappa\kappa}$, is the most widely used two-point statistic in weak lensing analyses. It quantifies the variance of two-dimensional density fluctuations as a function of angular scale $\ell$, and is defined by the ensemble average of the spherical harmonic coefficients of the field:

\begin{equation}
\left\langle \kappa_{\ell m} \kappa^*_{\ell' m'} \right\rangle = \delta_{\ell \ell'} \delta_{m m'} C_\ell^{\kappa\kappa},
\end{equation}

where $\kappa_{\ell m}$ are the spherical harmonic coefficients of the convergence field $\kappa(\boldsymbol{\theta})$. This power spectrum is directly related to the 3D matter power spectrum $P_\delta(k)$ through the Limber approximation and carries information about the amplitude and scale dependence of matter fluctuations. Cosmological parameters such as $S_8 = \sigma_8\sqrt{\frac{\Omega_m}{0.3}}$ can be constrained using measurements of $C_\ell^{\kappa\kappa}$.

\subsubsection*{Convergence Probability Distribution Function (PDF)}

The power spectrum contains Gaussian information of the convergence field, including its variance. To study higher-order moments, the one-point probability distribution function (PDF) of the $\kappa$ provides complementary information by encoding the full distribution of pixel values across the map containing information about the amplitude of the correlation functions of the field. The shape of the $\kappa$-PDF is sensitive to non-Gaussian features arising from nonlinear structure formation and includes contributions from higher-order moments like skewness and kurtosis \cite{Munshi2000,petri2016cosmology,Gatti2022}.

The $\kappa$-PDF is particularly effective at probing the tails of the convergence distribution, which correspond to rare structures such as massive clusters (high $\kappa$) or two-dimensional voids (low $\kappa$). PDF measurements recover information of the amplitude of one-point fluctuations and, once combined with information inferred from the power spectrum, provide improved constraints in the $S_8$ cosmological parameter \cite{Uhlemann2020,Thiele2023,Castiblanco2024}.%\jl{add Cora's PDF papers}

\subsubsection*{Minkowski Functionals}

Minkowski functionals (MFs) are a set of morphological descriptors that quantify the topology and geometry of scalar fields like $\kappa(\boldsymbol{\theta})$. In two dimensions, there are three functionals: the area fraction $V_0$, the perimeter length $V_1$, and the Euler characteristic (or genus) $V_2$, defined over excursion sets where the field exceeds a given threshold $\nu$.

Mathematically, for a field $\kappa$ and under a smoothing angular scale, the Minkowski functionals can be written as integrals over the excursion set:

\begin{align}
V_0(\nu) &= \frac{1}{A} \int_{\kappa > \nu} \mathrm{d}A, \\
V_1(\nu) &= \frac{1}{4A} \int_{\partial(\kappa > \nu)} \mathrm{d}l, \\
V_2(\nu) &= \frac{1}{2\pi A} \int_{\partial(\kappa > \nu)} \mathrm{d}l \, \kappa_c,
\end{align}

where $A$ is the total area of the field, $\partial(\kappa > \nu)$ denotes the boundary of the region where $\kappa$ exceeds $\nu$, and $\kappa_c$ is the local curvature along this boundary.

Minkowski functionals are sensitive to both Gaussian and non-Gaussian features in the field and have been shown to outperform standard two-point statistics in some cases, especially in the nonlinear regime \cite{kratochvil2012minkowski, munshi2012minkowski}. Their ability to capture the shape, connectivity, and filamentary structure of the convergence field makes them a powerful tool for extracting cosmological information beyond the power spectrum.

Combined constraints of MFs and $C_{\ell}^{\kappa\kappa}$ improve the constraints of $S_8$ in comparison to the power spectrum alone by 40\% ~\cite{Armijo2025}. 

%\vspace{0.3em}
Together, these summary statistics provide a multi-faceted description of the convergence field and are crucial for maximizing the scientific return of current and future weak lensing surveys. They are also instrumental in validating cosmological models and training data-driven methods such as machine learning emulators and likelihood-free inference techniques. \cite{Marques_2024,Cheng_2024,Novaes2024} %\newtext

\section{Simulations} \label{sec:simulations}

\begin{figure*}
    \centering
    \includegraphics[width=0.9\linewidth]{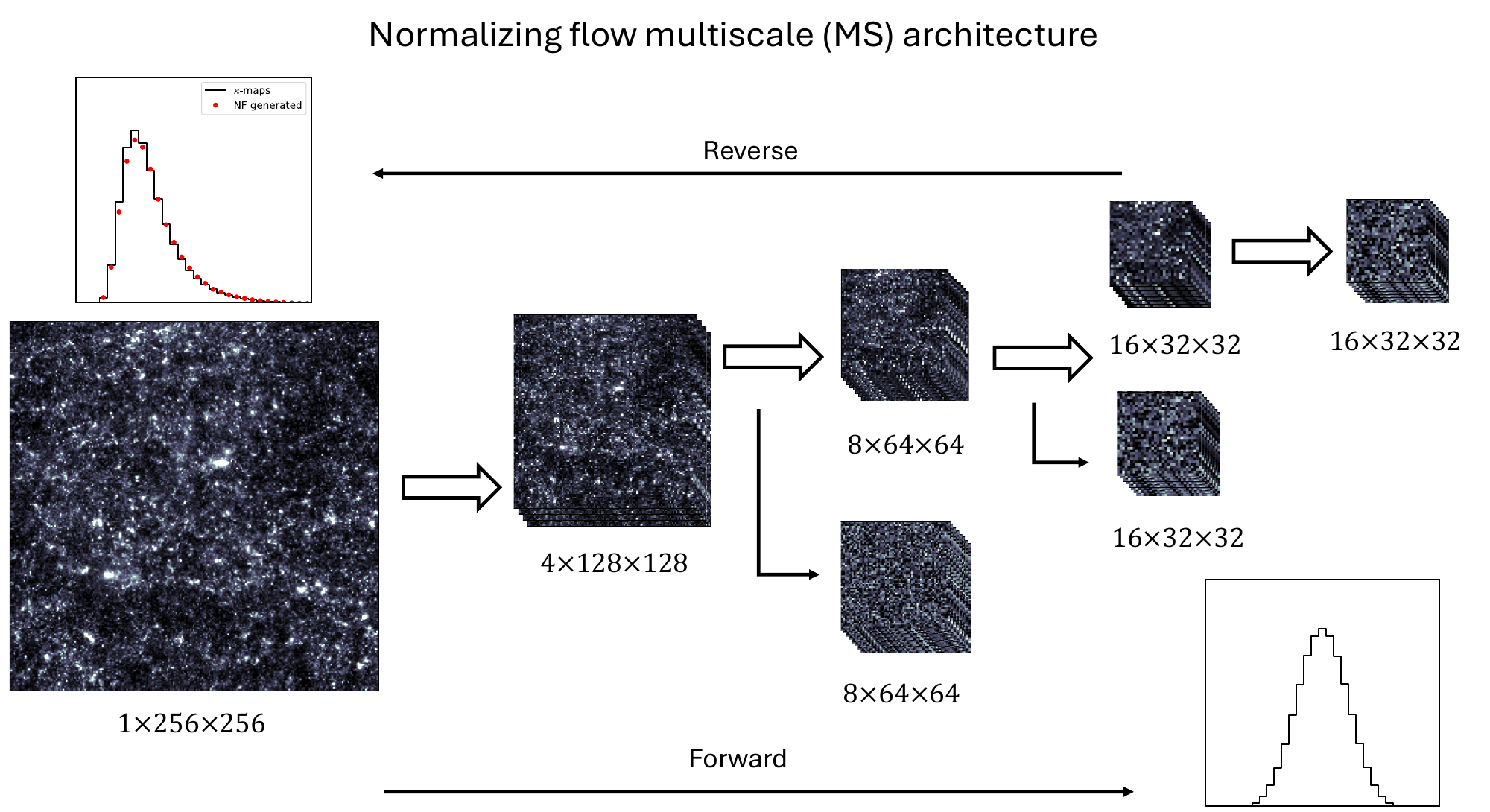}
    \caption{Diagram of multiscale NF network used for learning $\kappa$-map features. Images are passed through the flow with an initial size of $N_{\rm pix} = 256\times 256$ with a clear non-Gaussian distribution. After all the transformation, the latent space represents a Gaussian random field with the same image dimensions (number of pixels). The image representation is schemed in the center, showing how the original image, a convergence map, loses resolution by passing coupling, squeezing, and splitting transforms until it becomes indistinguishable from Gaussian random noise. The nature of the NF network requires the flow to be invertible, %\jl{I changed all ``reversible'' to invertible to be consistent}, 
    allowing for the generation of $\kappa$-maps from an initial Gaussian random field.}
    \label{fig:MS_Flow}
\end{figure*}

We model convergence field maps constructed from numerical simulations of galaxy cosmic shear. The SLICS simulations \citep{harnois2018slics} are a publicly available suite of high-resolution $N$-body simulations tailored specifically for weak lensing and large-scale structure analyses.

% \subsection{Convergence Maps from gravitational lenses}
To study weak lensing effects and interpret observational data, it is essential to produce convergence maps to study the statistical and physical properties of the Universe. One widely used approach is through the multiple-plane tiling technique~\cite{ValeWhite2003}. In the case of SLICS simulations, convergence is extracted using the Born approximation, where the deflection angles and shear tensors are computed from 18 mass sheets acting as the lens planes. \newtext{Throughout this work we use convergence maps constructed for sources in the lowest tomographic bin of a 5-bin LSST-like source-galaxy redshift distribution \citep{Zuntz2021}. The mean source redshift of this bin is $z_{\rm s} \approx 0.4$. We adopt this choice for two reasons. First, it isolates the regime in which non-Gaussianity is most pronounced, and as a more demanding test to the test the ability of the NF network to learn the joint distribution of pixel values and summary statistics. A generative model that performs well at low $z_{\rm s}$ is expected to perform at least as well at higher source redshift, where the field is closer to a Gaussian distribution.}

\newtext{SLICS simulations provide a sample of $N{\rm sims} = 954$ independent light-cones of $10\times 10$ deg$^2$ produced on a $7745\times 7745$ grid, used to calculate the covariance matrix for stage-III surveys, such as KiDS, DES, and HSC. Due to their high pixel resolution ($\sim 0.077$ arcmin) SLICS maps are also helpful for the forecasting of stage-IV surveys, like Euclid and LSST. We will use these maps as training data for our machine learning model to test how well the summary statistic information can be learned. For computational performance and to match the typical resolution at which weak-lensing summary statistics are evaluated in practice, we downgrade each map to a common working resolution before analysis. The exact multiple explored resolutions used in each test are specified in Section V.}

\section{Flow based neural networks} \label{sec:NF}

Normalizing flow (NF) networks constitute a class of generative models that learn complex probability distributions by transforming a simple base distribution, such as a multivariate Gaussian, through a sequence of invertible and differentiable mappings. By defining a series of these bijective transformations, a flow maps a latent distribution $p(z)$ into a target data distribution $p(x)$ according to the change-of-variables rule. Each transformation is designed to be invertible, with a tractable Jacobian determinant, which allows exact likelihood evaluation and efficient sampling. Flow-based models can be implemented using various architectures, with defined transformations, offering a trade-off between expressiveness and computational efficiency. In contrast to generative adversarial networks (GANs) and variational autoencoders (VAEs), normalizing flows maintain an exact, one-to-one correspondence between the latent and data spaces, preserving dimensionality throughout the training process. In our task to replicate information from weak lensing $\kappa$ fields, we focus in two types of architectures: Neural spline flows and multiscale flows. 

%\jl{for each subsection, can you write out the actual function used in your network for NF? even just a rough form, to give readers some idea of what they look like? also the equation for the loss function. did you use KL divergence for both NSF and MS?}

%$\bullet$ Explain Normalizing flows.\\

\subsection{Neural spline flow}
Neural spline flow (NSF) networks are a class of normalizing flow models that employ a collection of polynomial splines to construct transformations for density estimation. By using rational--quadratic or higher-order spline functions, NSF enables flexible and invertible mappings between simple base Gaussian distribution and complex target distributions~\citep{Durkan2019}. The splines are parameterized by neural networks, which predict the knot positions and derivatives, providing a data-dependent transformation at each layer. Combining the adaptability of splines, the NSF is helpful for the generative sampling and probabilistic modeling. For example, for learning the summary statistic of a set of simulations, we employ the NSF to replicate the likelihood distributions of weak-lensing summary statistics. The networks used to learn these target distributions are relatively compact, containing $N_{\rm par} \simeq 3\text{\textendash}4\times10^5$ trainable parameters.  \newtext{For our implementation, each model consists of five masked autoregressive flow (MAF), where each block defines collection of splines with a defined Jacobian determinant~\citep{Papamakarios2017}.}

%$\bullet$ Explain the networks used in this work

\subsection{Multi-scale normalizing flow}

\begin{figure*}
    \centering
    \begin{tabular}{ccr}
      \includegraphics[width=0.3\linewidth]{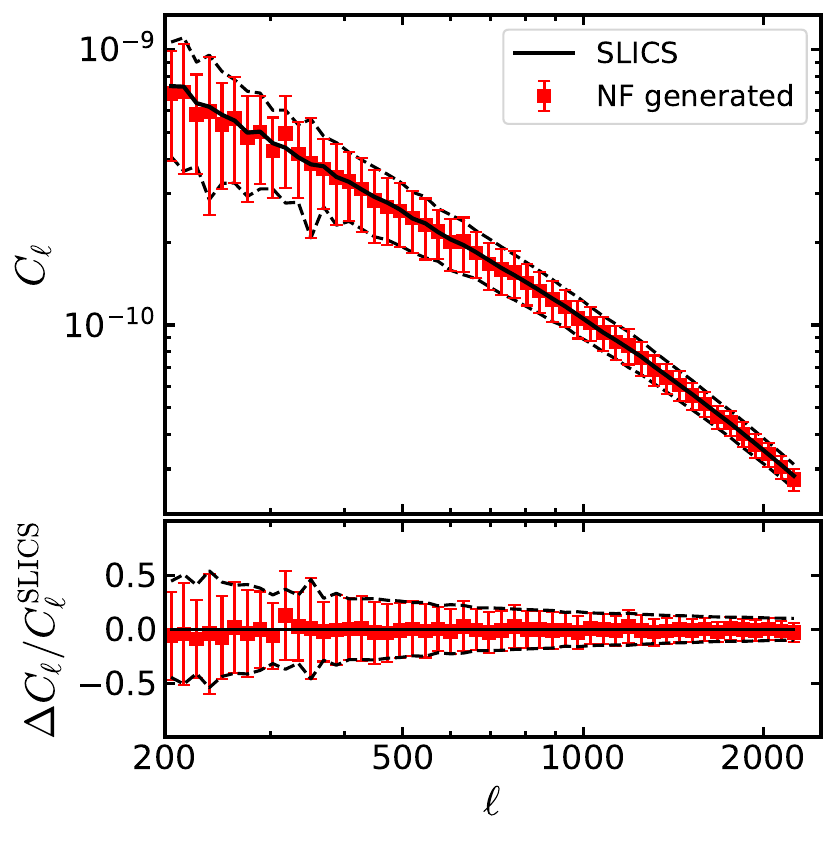} & 
      \includegraphics[width=0.3\linewidth]{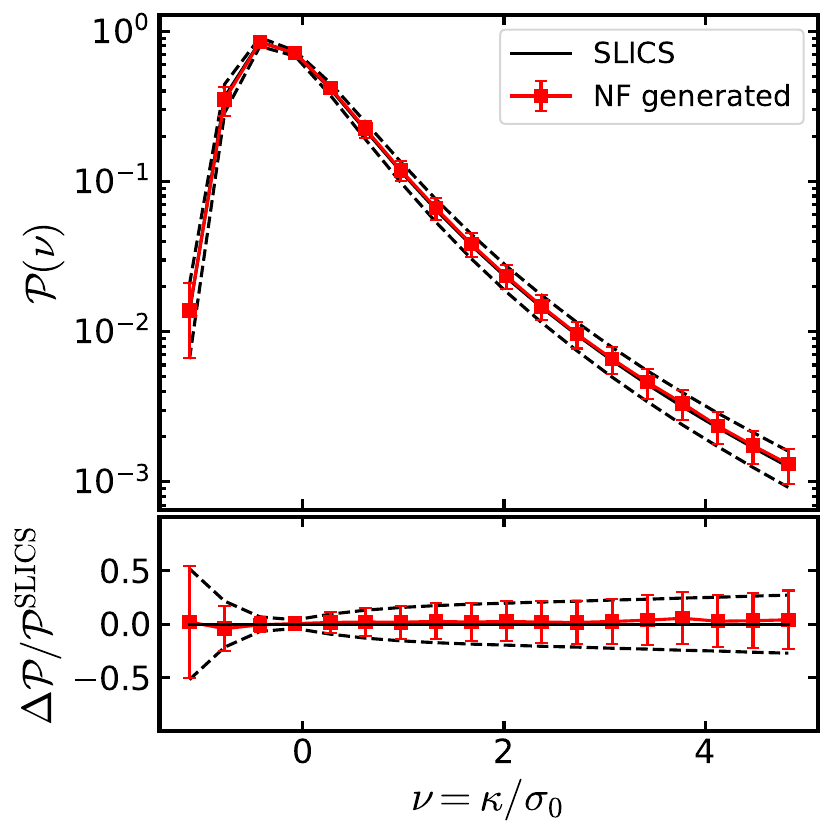} & \\
      \includegraphics[width=0.3\linewidth]{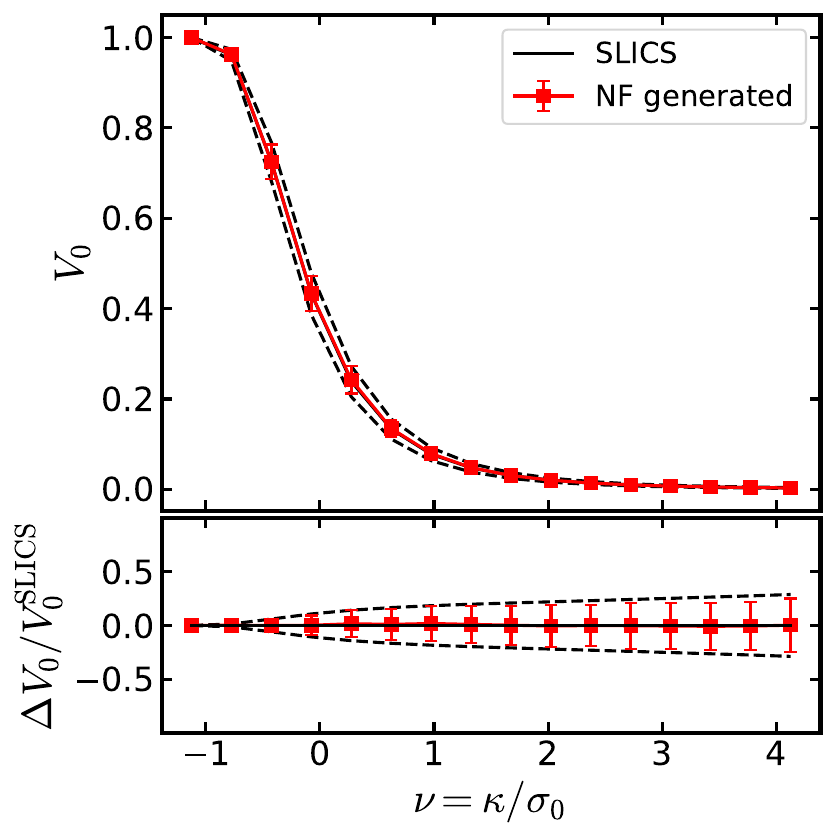} & 
      \includegraphics[width=0.3\linewidth]{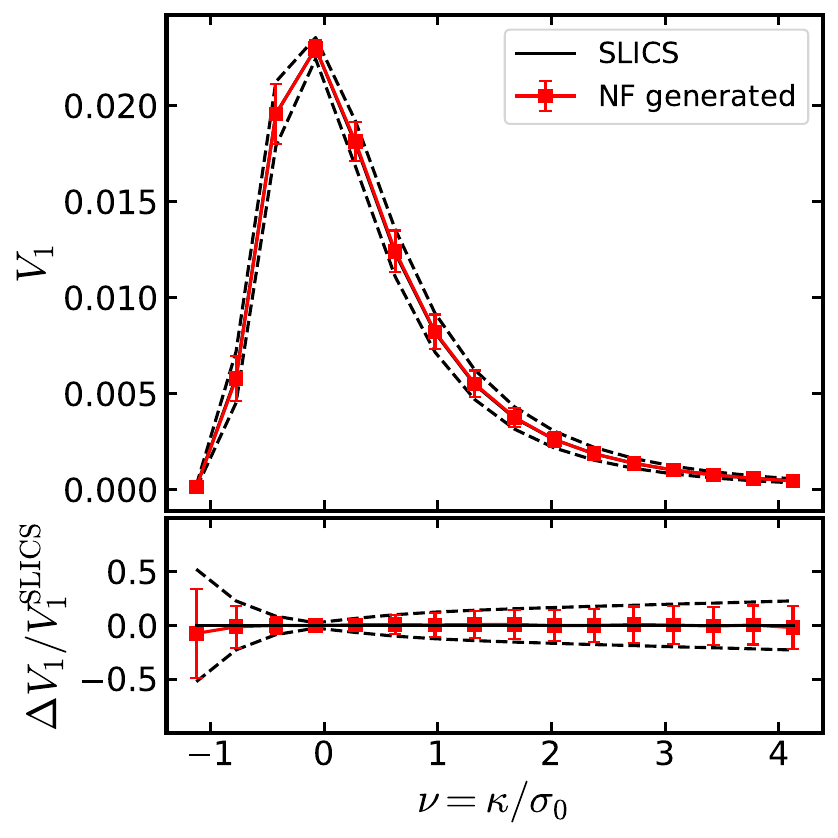} &
      \includegraphics[width=0.3\linewidth]{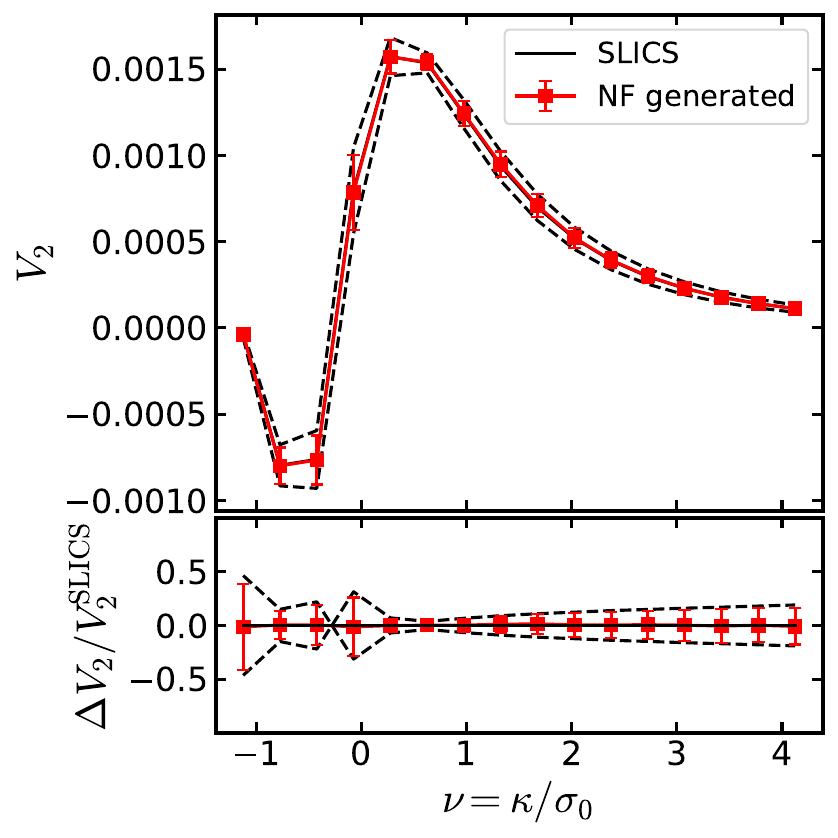} 
    \end{tabular}
    
    \caption{Replication of summary statistics using NF network and SLICS simulations (ground truth) dataset. We calculate the mean of the angular power spectrum $C_{\ell}^{\kappa\kappa}$, $\kappa$-PDF (top panels) and three Minkowski functionals $V_0$, $V_1$ and $V_2$ (bottom panels) for the SLICS simulation (black lines) \newtext{and the synthetic data vectors generated using the Neural Spline Flow} 
    %\jl{in caption, write out the full name, as NSF means nothing for people only read captions} 
    network. The bottom sub-panels include the relative difference compared to the ground truth data vector. We add the standard deviation of SLICS (black dashed lines)%\jl{sample here you meant slics? then write out slics} 
    and the synthetic data (red error bars).}%\jl{the caption for SLICS is inconsistent in top left panel}}
    \label{fig:DV_stats}
    
\end{figure*}

%$\bullet$ Network for generative models, a powerful, interpretable and versatil stochastic model. \\
Multiscale architectures in normalizing flows are designed to efficiently model the complex, hierarchical structure of image data by progressively transforming data at multiple spatial resolutions. Instead of processing the entire image at full resolution throughout the flow, these architectures split the data into different scales, typically by factoring out low-level details at earlier stages and modeling high-level structure in coarser representations~\cite{Dinh2016, Kingma2018}. This strategy not only reduces computational costs but also improves modeling flexibility and sample quality, making it particularly effective for image generation tasks where capturing both global structure and fine-grained details is crucial. This particular architecture becomes relevant when studying $\kappa$-maps, as these follow the hierarchical large-scale structure, making it crucial for studying the convergence field. In practice, this is the architecture used to replicate the information from the $\kappa$-maps and relies on different multiscale levels to learn the non-Gaussian information. In Figure~\ref{fig:MS_Flow} we describe this architecture, highlighting the different multiscale transforms. These are based on several invertible coupling layers, squeezing, and splitting operations defined in \cite{Dinh2016} to create a multiscale flow that generates the target distribution from pure Gaussian noise. In total, several squeeze layers and split layers for maps with sizes of $64 \times 64$ , $128 \times 128$ and, $256 \times 256$ pixels (one color Chanel), resulting in several resolution map representations. When applying the coupling layers, a convolutional neural network (CNN) is used to learn the coefficients of the affine transformation. After all the transformation have been applied, the original image is taken to a latent space representing pure Gaussian noise. Once trained, the reverse process starts from the noise field at the same resolution, turning to a confident $\kappa$-map with the same pixel distribution as the original maps. To train flow-based models, the loss function is based on the Kullback-Leibler (KL) divergence, which serves to minimize the distance between the distribution modeled by the flow and the proposed (Gaussian) prior. For the objective of learning the pixel distribution of convergence maps, we study multiscale architectures with different sizes: A simple architecture is composed of 518K parameters (MS1), and adding additional multiscale layers results in 665K parameters (MS2). For larger images ($256 \times 256$), we use a multiscale architecture with 893K parameters (MS3).
%We summarize the networks used for both learning the likelihood of the summary statistics based in NSF and the multiscales architectures for $\kappa$-map generations in Table \ref{tab:model_sizes}. 

\begin{figure*}
    \centering
    \begin{tabular}{ccr}
        \includegraphics[width=0.32\linewidth]{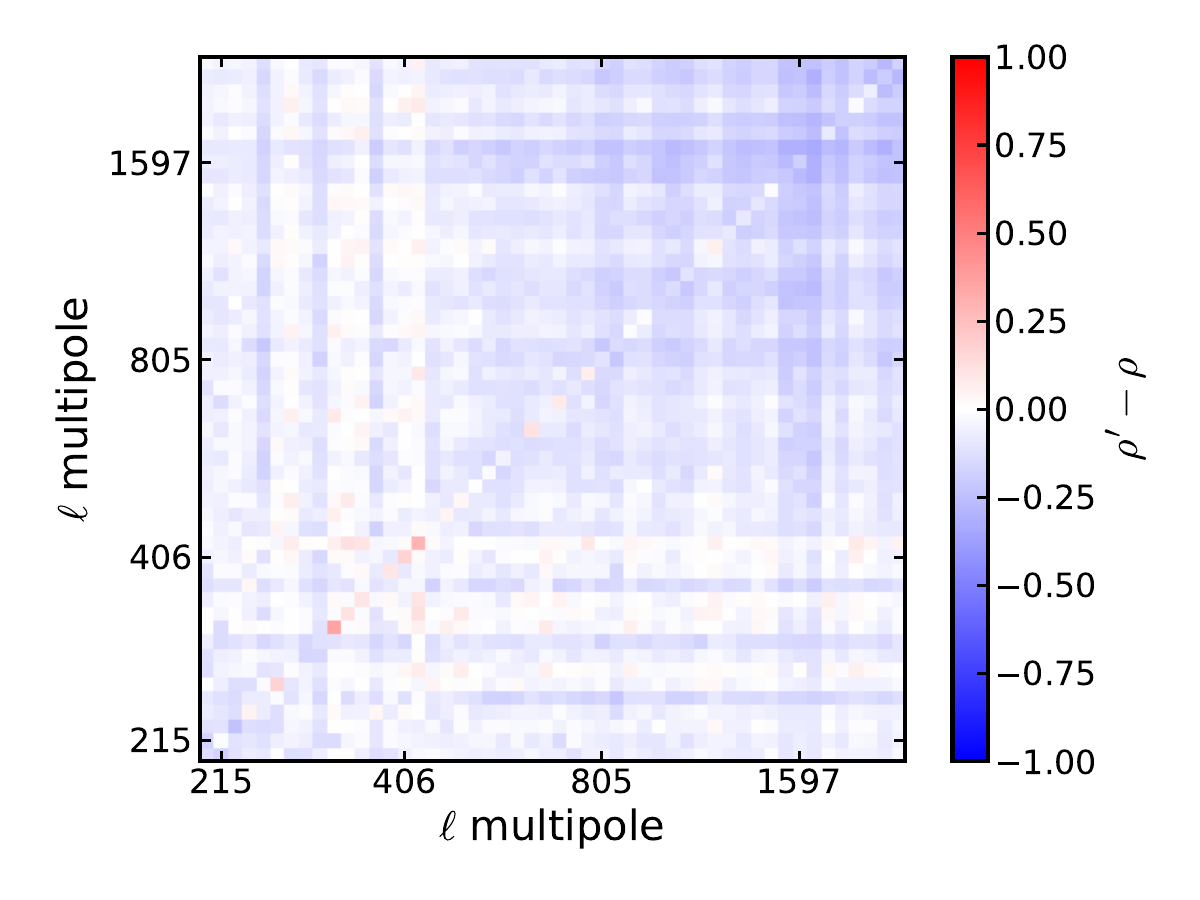} & 
        \includegraphics[width=0.32\linewidth]{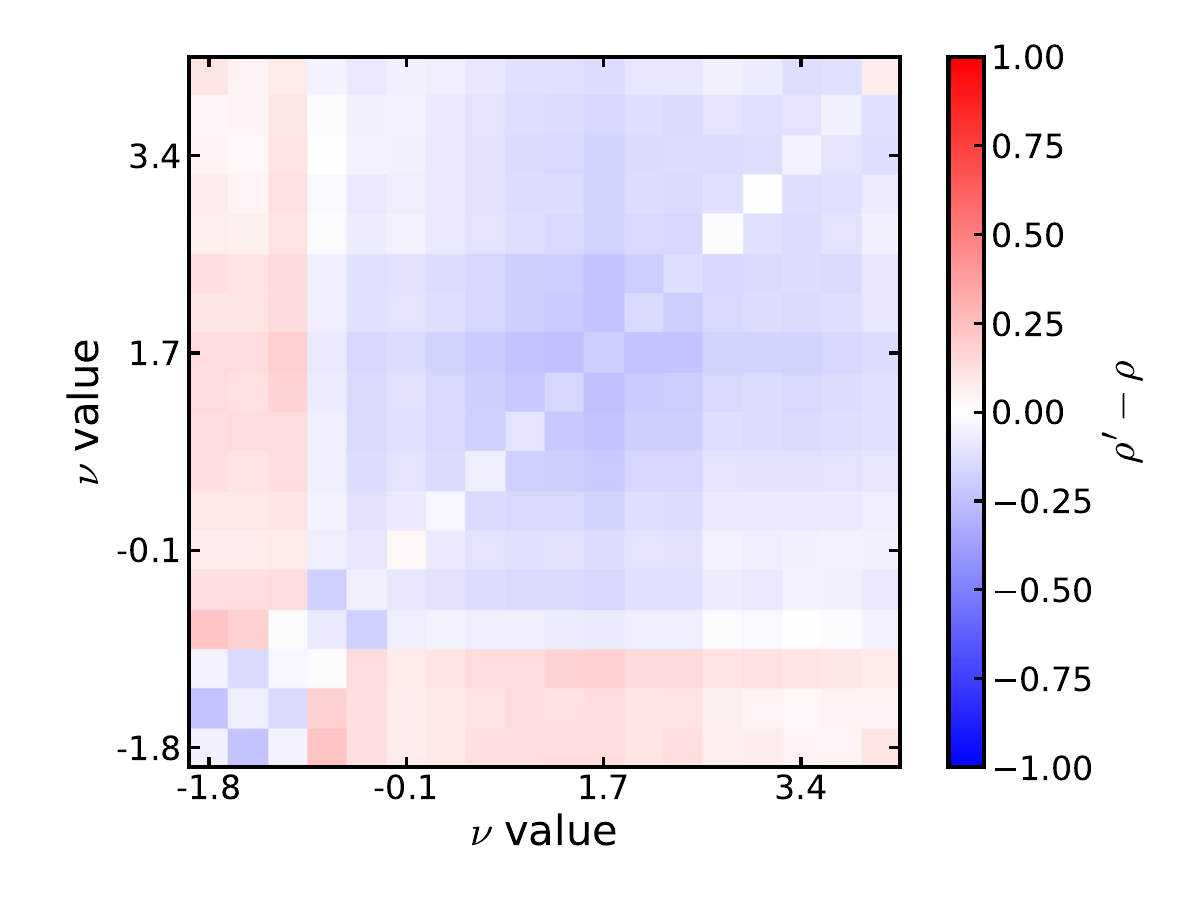} &
        \includegraphics[width=0.32\linewidth]{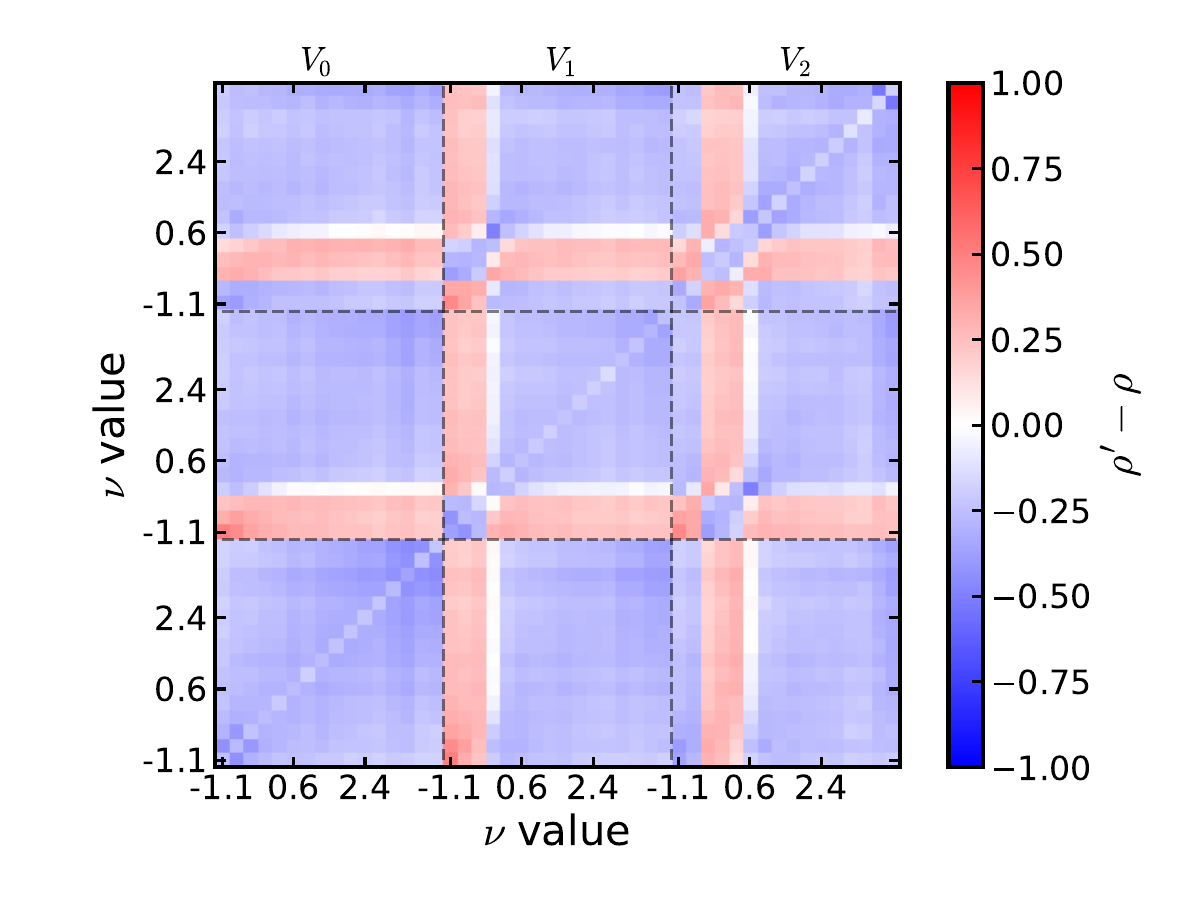}
    \end{tabular}

    \caption{residual of correlation coefficients $\rho^{\prime} - \rho$ of the covariance matrices of the summary statistic showed in Fig. \ref{fig:DV_stats}, $C_{\ell}^{\kappa\kappa}$ (left), $\kappa$-PDF (middle), and Minkowski functionals (right) for NF generated data vectors and SLIC simulations. %\jl{The Normalizing Flow models appear to underestimate the covariances} \ja{In general yes, but is low and only when I use large data vectors, If I average is <10\%}.%Coefficients are colored using the value coefficient values $\sqrt{C_{ij}^{\rm NF}/C_{ij}^{\rm Sim} - 1}$, indicating how well the values agree with the ground truth.
    %\jl{the diagonal term seems close to 0 here, but a lot bigger in fig.4 to me.. like 50\%??}\ja{Yes, in fig. 4 my claim that this is catastrophic for maps if the network is not modified.}
    }
    \label{fig:DV_stats_cov}
\end{figure*}

%\begin{table}[]
%    \centering
%    \begin{tabular}{c|c|c|c|c}
%    \toprule
%       & NSF  & MS1 & MS2 & MS3  \\
%       \midrule
%       $N_{\rm par}$ & $410870$ & $518170$ & $664500$ & $893004$ \\
%    \bottomrule
%    \end{tabular}
%    \caption{Caption \newtext{Need to add sizes...}}
%    \label{tab:model_sizes}
%\end{table}

\section{Results}\label{sec:results}

\subsection{Replicating $C_{\ell}$ and non-Gaussian statistics}
\newtext{We test the recovery of the summary statistics as a quality diagnostic of the NF-generated samples: a model that accurately reproduces the mean, variance but fails on off-diagonal terms has not learned the full joint distribution, and would therefore be unreliable as an emulator for cosmological inference.}

We first examine the replication of summary statistics using the NSF network trained on the individual data vectors corresponding to the different summary statistics. Figure~\ref{fig:DV_stats} presents a comparison between the angular power spectrum $C_{\ell}^{\kappa\kappa}$, the $\kappa$-PDF, and the Minkowski functionals (MFs) computed from the SLICS simulations and those reproduced by the trained NSF model. \newtext{The summary statistics in this section ($C_{\ell}^{\kappa\kappa}$, $\kappa$-PDF and Minkowski functionals) are computed from the SLICS maps downgraded to $256\times 256$ pixels over the full $10\times 10$ sq. degree field. The PDF and Minkowski functionals are evaluated on maps smoothed with a Gaussian kernel of $\theta_{\rm s} = 2$ arcmin. FWHM, while the Power Spectrum is computed on the unsmoothed (downgraded) maps.}. \newtext{The NSF is trained on the $N_{\rm sim} = 954$ summary-statistic data vectors computed from the SLICS realizations. After training, $N_{\rm gen} = 1000$ synthetic data vectors are drawn from the NF and used to evaluate the mean, variance, and covariance shown in Figures \ref{fig:DV_stats} and \ref{fig:DV_stats_cov}.}  The results show a close agreement with the ground truth, with differences typically below the 1\% level across different binning schemes in $\ell$-multipoles and normalized field values $\nu$. Both the mean and standard deviation are shown in Figure~\ref{fig:DV_stats}, demonstrating that the network successfully learns these statistical properties directly from the binned distributions of each summary statistic.

We show in Figure~\ref{fig:DV_stats_cov} the ratio of covariances between the NF-generated and simulated datasets for each summary statistic. In the case of the covariance comparison, we focus primarily on the off-diagonal components and compute their relative values as \newtext{$\rho^{\prime} = C_{ij}^{\rm NF}/\sqrt{C_{ii}^{\rm Sim}C_{jj}^{\rm Sim}}$, where $C_{ij}^{\mathrm{NF}}$ denotes the covariance of the summary statistic generated by the NF network, and $C_{ij}^{\mathrm{Sim}}$ represents the corresponding covariance from the SLICS simulations. We adopt this definition $C_{ii}^{\mathrm{Sim}}$ instead of the classical correlation coefficient, $\rho = C_{ij}/\sqrt{C_{ii}C_{jj}}$, for comparing directly with the diagonal terms of the SLICS covariance (the variance at $i$ and $j$ bins). This choice facilitates the interpretation of fractional differences in the covariance structure comparing directly with the baseline covariance provided by simulations. Therefore in the ideal test when the covariance matrix is fully recovered by generated sample $\rho^{\prime} - \rho$ is close to zero.}

Figure~\ref{fig:DV_stats_cov} shows that the off-diagonal terms of the covariance are underestimated by the network by approximately $5\%$ for $C_{\ell}^{\kappa\kappa}$, $5\%$ for the PDF, and $13\%$ for MFs, as indicated by the light-blue colors. Some of these values might be expected, as some of the statistics are divided into several data bins, such as $C_{\ell}^{\kappa\kappa}$, making the covariance be more sensitive to sampling noise. Overall, these values remain stable across the different binnings of $\ell$-multipoles and normalized field values $\nu$ tested for the data vectors. The accuracy, however, decreases slightly for configurations with $N_{\ell\text{-bins}}>50$ and $N_{\nu\text{-bins}}>20$. \newtext{However, there is a clear tendency to underestimate the off-diagonal covariance terms (blue gradient) for higher multipoles $\ell > 805$ when compared with the variance of SLICS simulations. This is repeated in the NG statistics, such as the values for MFs, which are poorly recovered in the generated sample.}

\begin{figure*}[!htbp]
    \centering
    \includegraphics[width=0.5\linewidth]{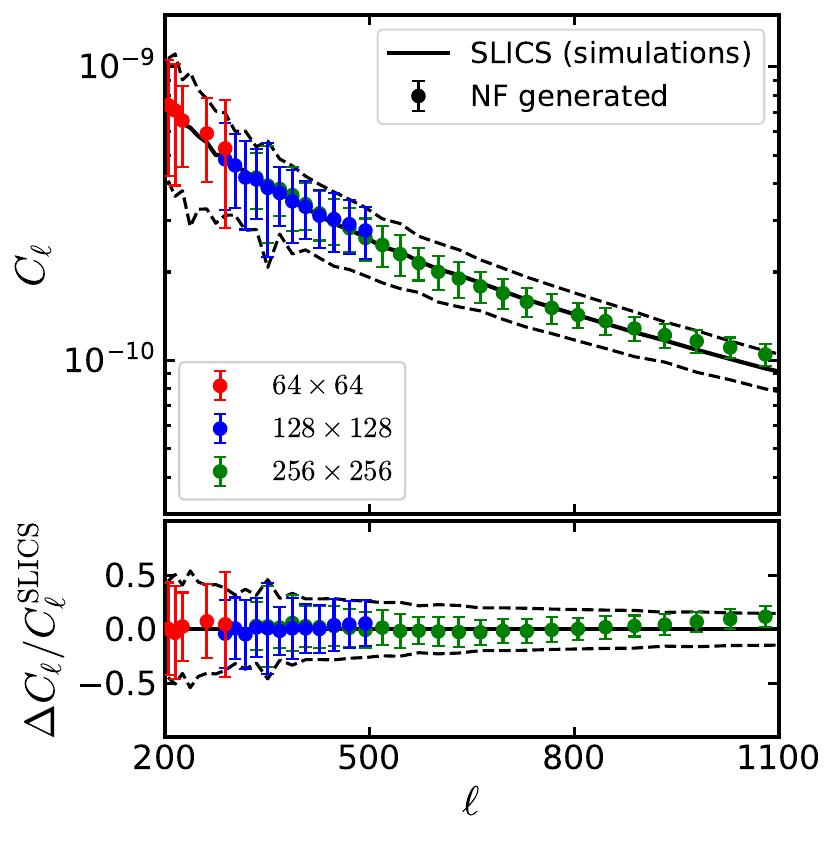}
    \caption{Angular Power spectrum $C_{\ell}^{\kappa\kappa}$ of NF generated convergence maps compared to the $\kappa$-maps from SLICS simulations. These are calculated from maps with different pixel sizes: $64\times 64$ (red), $128\times 128$ (blue), and $256\times 256$ green. We add the relative difference w.r.t. the ground truth in the bottom panel.%\jl{the variance seems to be quite underestimated by NF, by 50\%?}\ja{Indeed, at this level I'm not claiming to recover the variance completely, and there's the tendency to decrease with dimensionality.}
    }
    \label{fig:C_ell_maps_resol}
\end{figure*}

\begin{figure*}
    \centering
    \begin{tabular}{ccr}
        \includegraphics[width=0.3\linewidth]{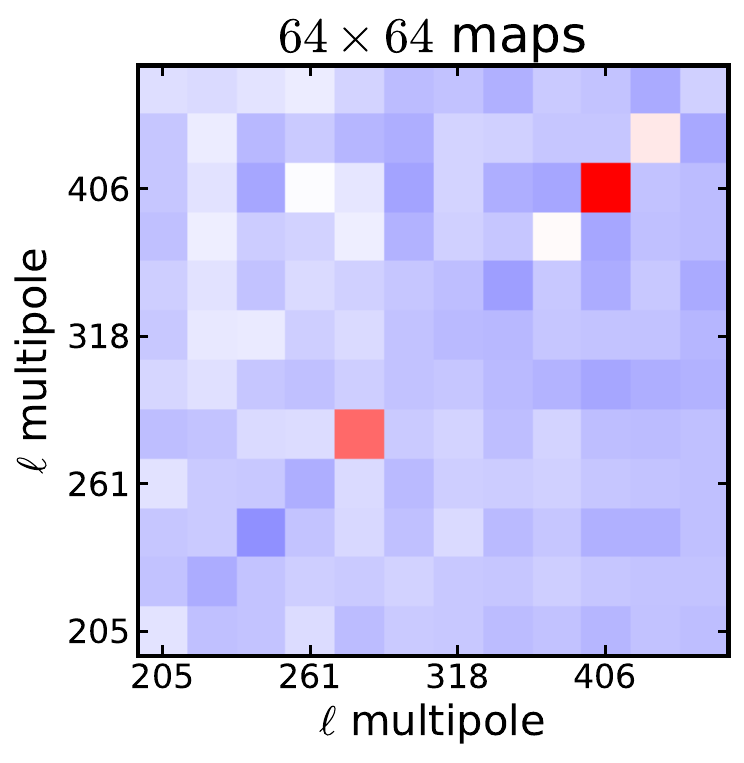} & 
        \includegraphics[width=0.282\linewidth]{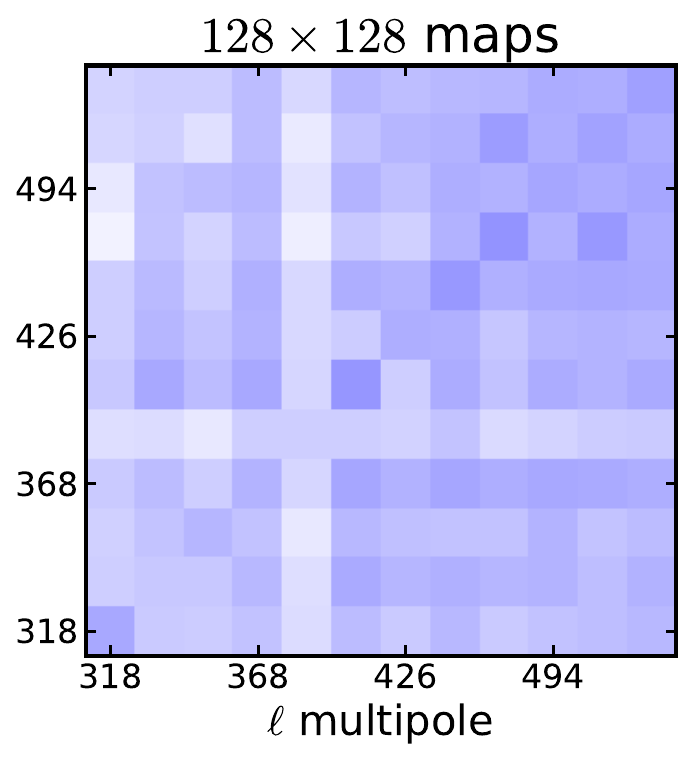} &
        \includegraphics[width=0.378\linewidth]{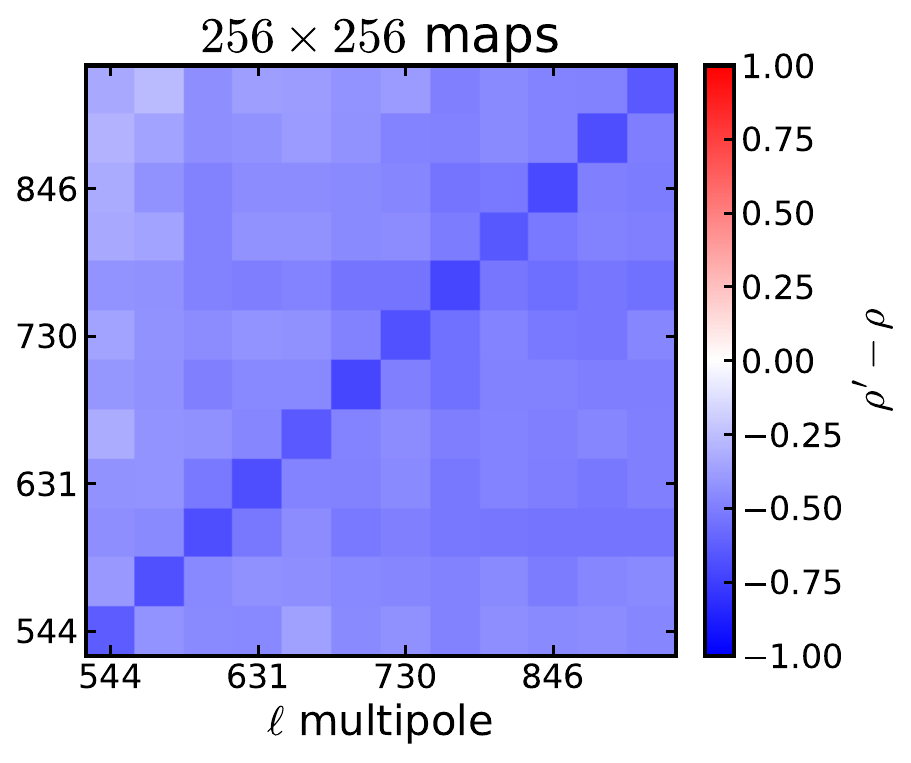}
    \end{tabular}
    \caption{Residual $\rho^{\prime} - \rho$ of covariance matrices for $C_{\ell}^{\kappa\kappa}$ in Fig. \ref{fig:C_ell_maps_resol}. We select 12 bins for each resolution map to help the comparison between the cases:  $64\times 64$ (left), $128\times 128$ (middle), and $256\times 256$ right.}%\jl{instead number 0-12, label the actual ell range}}
    \label{fig:cov_maps_resol}

\end{figure*}

\subsection{Statistic of NF generated $\kappa$-maps}

%$\bullet$ Show maps, show multiscale, show counts histograms, and maybe transformations.\\
The next step is to test whether, beyond reproducing the information contained in the summary statistics of the simulations, NF networks can also replicate complete $\kappa$-maps by learning the distribution of individual pixels. In this case, square images of size $N \times N$ pixels are passed through the flow and transformed into Gaussian noise images of identical dimensions. Once trained, the inverse flow generates new realizations of the convergence field from any initial Gaussian random field. As the dimensionality of the problem increases rapidly with the image size (scaling as $N^2$), in contrast to the previous case based on one-dimensional data vectors, we adopt the MS flow architecture. This design improves the performance by introducing convolution-like transformations that efficiently capture spatial correlations within the maps. To evaluate how the network results vary with image size, we test three resolutions. Figure~\ref{fig:C_ell_maps_resol} presents results for three different map sizes: $N_{\mathrm{pix,1}} = 64 \times 64$, $N_{\mathrm{pix,2}} = 128 \times 128$, and $N_{\mathrm{pix,3}} = 256 \times 256$. Because the different resolutions probe distinct angular scales, they correspond to different ranges in angular scale, represented by a range of $\ell$ multipoles. We limit the comparison of the replicated power spectrum to the range $200 < \ell < 1100$, determined by the image resolution and map size.

\begin{figure*}
    \centering
    \begin{tabular}{cccr}
    \includegraphics[width=0.3\linewidth]{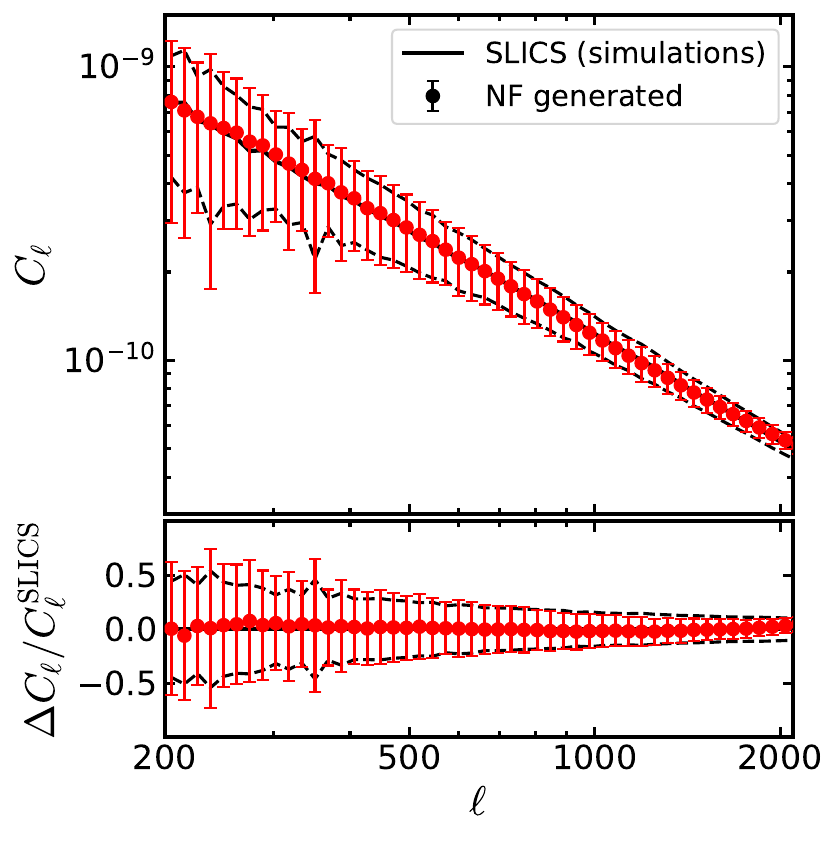} &
    \includegraphics[width=0.3\linewidth]{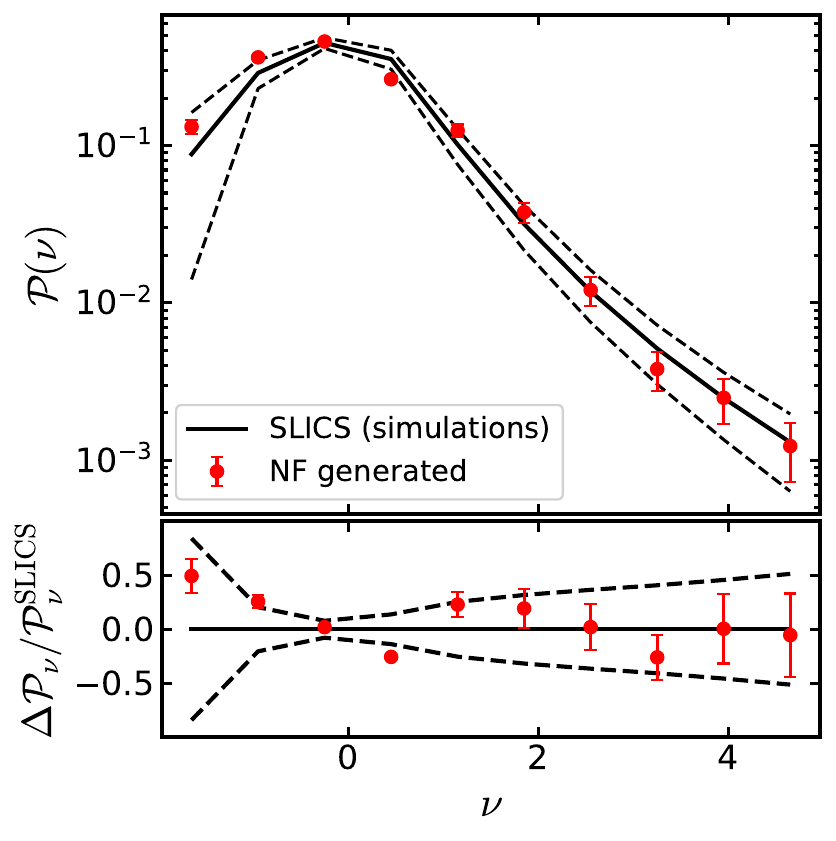} & \\
    \includegraphics[width=0.3\linewidth]{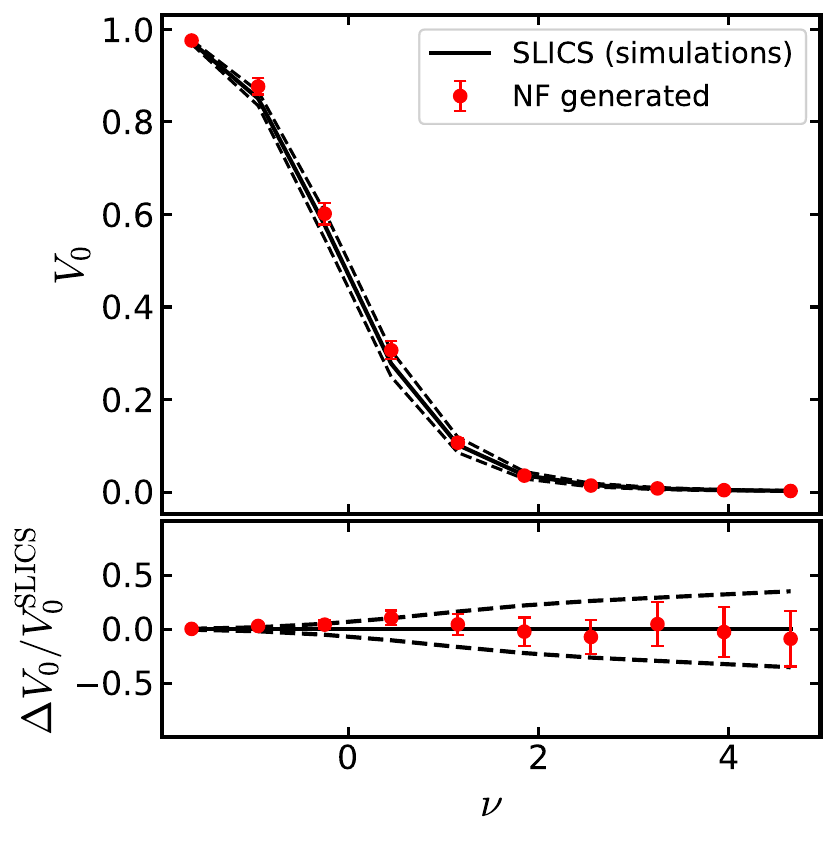} &
    \includegraphics[width=0.3\linewidth]{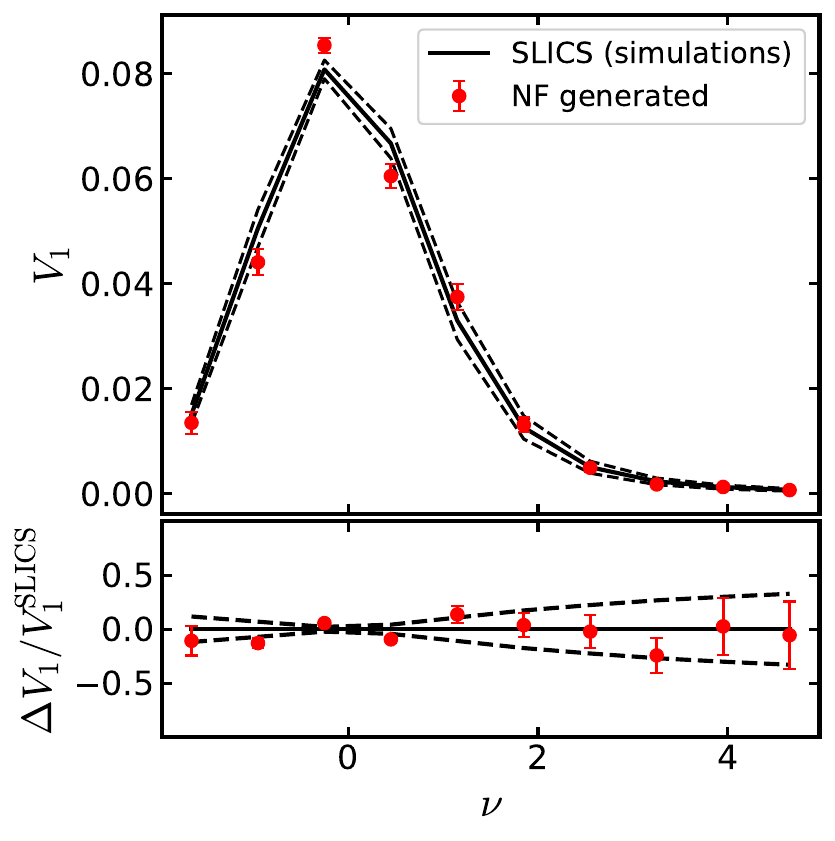} &
    \includegraphics[width=0.3\linewidth]{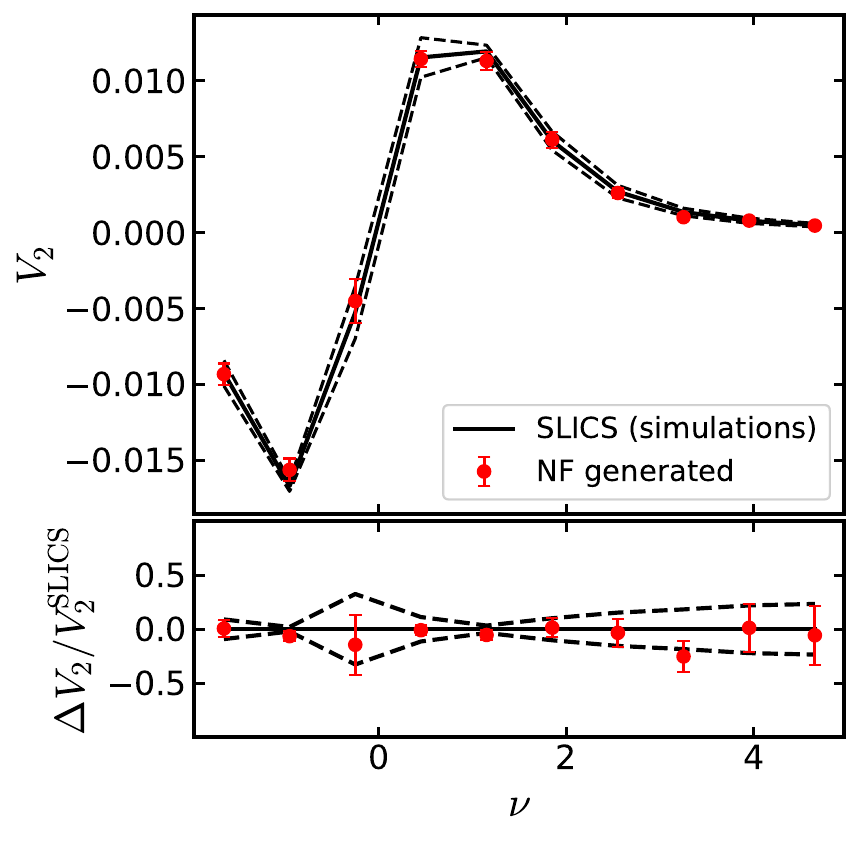}
    \end{tabular}
    \caption{Same as Figure~\ref{fig:DV_stats} but with NF generated maps using the multiscale architecture. We calculate the summary statistics using $\kappa$-maps with included pixel noise $\sigma_{\kappa}$.}%\jl{what's going on with the NF error bars? they're so severely underestimated.. but the jumping data points seem to reflect the error more correctly; also this is with pixel noise? if so, describe in the caption. why don't you use the original map without augmentation or noise} \ja{Right, this is for noisy data, which gives the best results for $C_{\ell}$, so I thought it would be similar for PDF, MFs. It doesn't change too much with the other cases, i remember this one was the best one by eye.}}
    \label{fig:maps_stats}
\end{figure*}

For the selected range, the mean of the power spectrum $C_{\ell}^{\kappa\kappa}$ is recovered with percent-level accuracy, %\jl{write out the actual \%, 1\%?}, 
within approximately $3\%$ of the ground truth estimated from the SLICS simulations and consistent within one standard deviation. However, the standard deviation of the power spectrum is systematically underestimated for the NF-generated maps. This underestimation is also evident in the covariance matrices of the power spectrum shown in Figure~\ref{fig:cov_maps_resol}, where an excess of blue regions indicates lower values relative to the simulations. The accuracy with which the covariance terms are recovered $\rho^{\prime}$ decreases with increasing map size, becoming most pronounced for $N_{\mathrm{pix,3}}$, where even the variance (diagonal terms) is at least $25\%$ smaller than the ground truth. \newtext{We attribute this loss of accuracy to the rapid increase in dimensionality with image size, when using the MS1 model only. This trend is clearly observed as both the variance and covariance accuracy degrade with higher dimensionality, reaching discrepancies up to $75\%$ for the largest maps ($N_{\mathrm{pix,3}}$).}

%\jl{is there a reason why MS map stats are so much worse than trained on NSF? is it because maps are just more difficult to train? what if you force the stats into the loss function? did you try?}

% \subsection{Increasing the accuracy of the variance and covariance of NF generated maps}
\subsection{Network size, data augmentation, and noise}

\begin{figure*}
    \centering
    \includegraphics[width=0.95\linewidth]{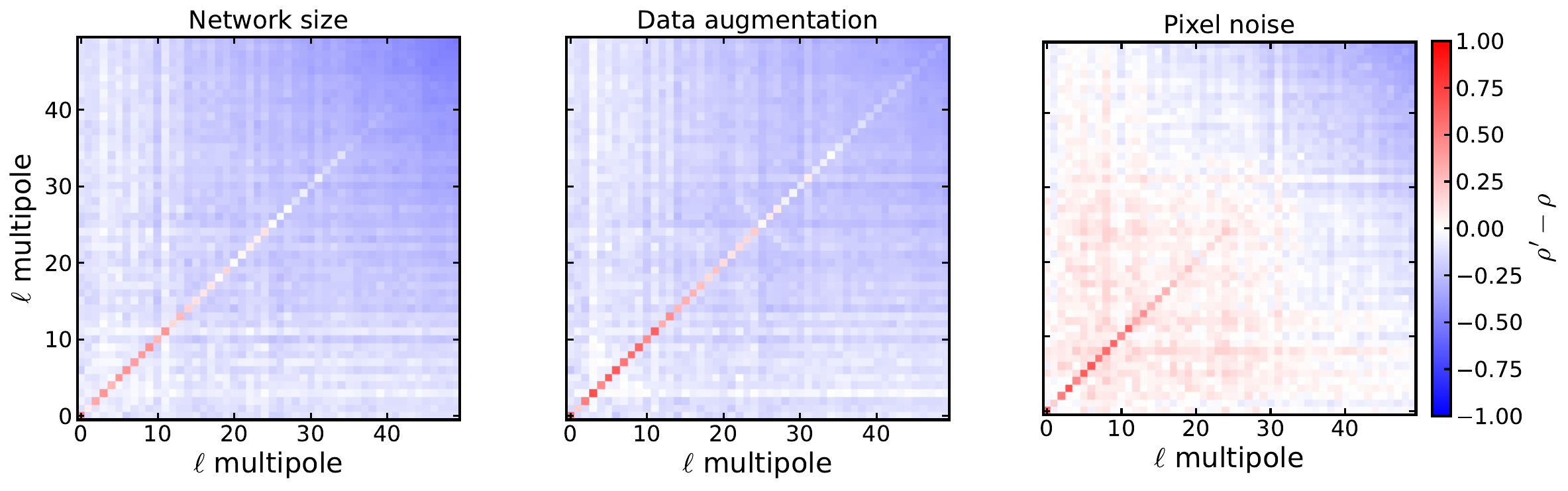}
    \caption{$C_{\ell}^{\kappa\kappa}$ covariance of NF generated maps for different training cases: Using a larger network size for training with $\kappa$-maps (left), using data augmentation of the sample (middle), and adding pixel noise to the samples when training (right). The color bar coding is the same as in Figure~\ref{fig:DV_stats_cov}}.%\jl{write out the actual ell range instead of bins, as bin numbers carry very little info; is this clkk? if so, include in caption}}
    \label{fig:cov_Cell_maps}
\end{figure*}

To assess whether the NF-generated maps can reproduce the same statistical properties as those computed from the simulations, we evaluate how well the network architecture handles the increased dimensionality of the $\kappa$-map images. We explore three strategies: increasing the network size, applying data augmentation during training, and adding stochastic noise to the training samples.

First, we extend the depth of the multiscale (MS) flow architecture described in Section~\ref{sec:NF}. Specifically, we introduce two additional levels of squeeze-and-split operations within the flow, which enable the network to better capture multi-scale structure in the input maps. This approach has been shown by \citet{Dai_Seljak2023} to enhance the performance of flow-based models in replicating weak-lensing images. Our implementation includes three multiscale levels in total, with additional coupling layers inserted at each level following the RealNVP architecture of \citet{Dinh2016}. This design allows the network to efficiently learn the spatial correlations across multiple resolutions.
%moved checkerboard and channel explanation to the methods part.

We also implement a data augmentation framework to make more efficient use of the limited number of simulated $\kappa$-maps available for training. Following a procedure similar to that of \citet{Ribli2019}, we modify the original images to expand the effective training set. Starting from the $10\times10~{\rm deg}^2$ SLICS maps, we randomly select a subset of maps and crop square regions of $5\times5~{\rm deg}^2$, which are then randomly rotated by $90^\circ$ or $180^\circ$. This process increases the number of training samples without altering the statistical properties of the replicated summary statistics, which are rotationally invariant. However, it helps the network to include more images during the training process. It also exposes the network to a broader variety of local features during training. The selection of this subsample is done in order to artificially increase the variance of the training set. Although this may slightly change the target power spectrum, as we remove the large modes outside the $5~{\rm deg}$ window, previous studies \citep{Pope2008,Petri2016,PearsonSamushia2016} have shown that the mean of the summary statistics can be accurately captured using only a fraction of the original sample. We maintain the same ranges of multipole scales and normalized field values $\nu$, as these are confidently represented by the summary statistics derived from both the $10~{\rm deg}$ and $5~{\rm deg}$ aperture maps. In addition, we generate a complementary dataset of noisy images by adding Gaussian noise with a standard deviation of $\sigma_{\kappa} = 0.008$, chosen to approximate an LSST-like survey. We treat this as an independent augmentation case, since the addition of noise is expected to affect the summary statistics while also acting as an effective regularization technique during training~\citep{Shirasaki2019,zhong2024}.

\begin{figure*}
    \centering
    \includegraphics[width=0.90\linewidth]{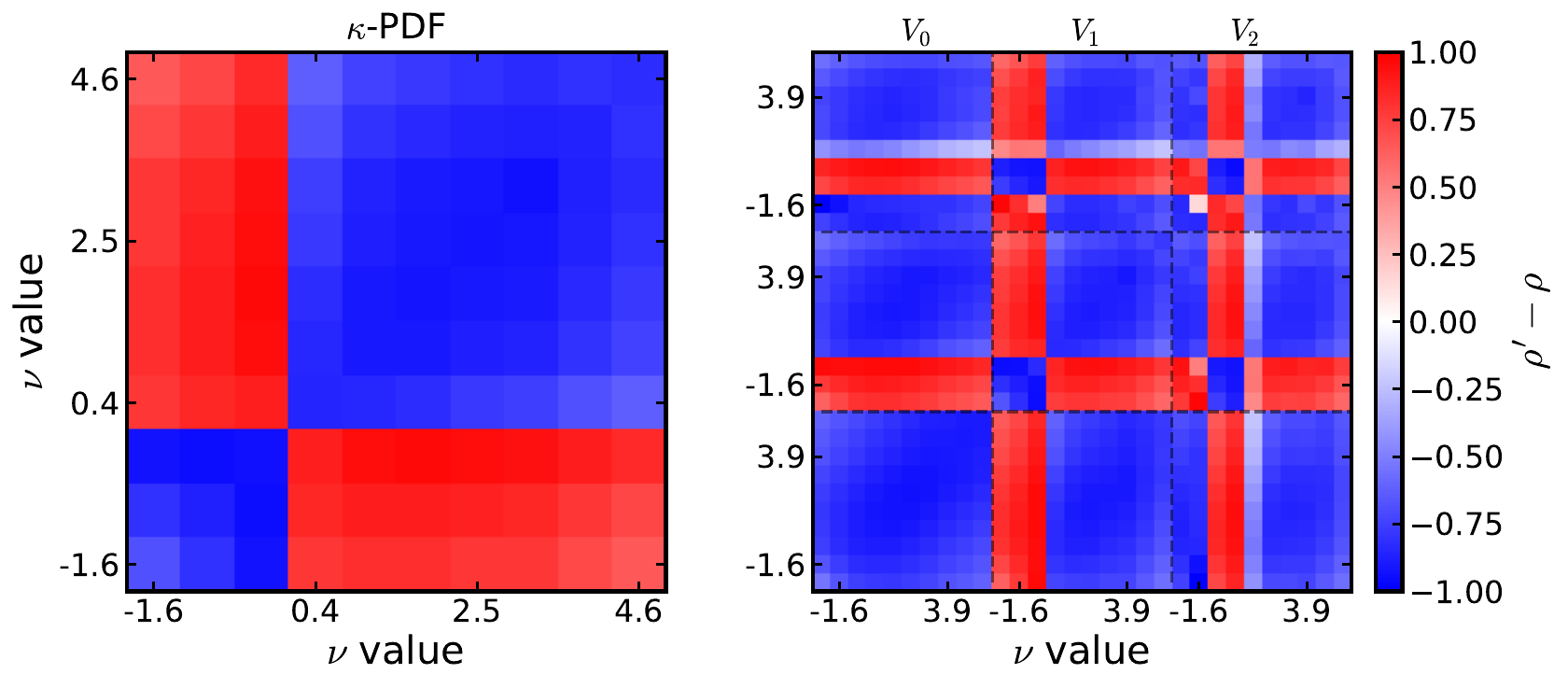}
    \caption{Same as in \ref{fig:cov_Cell_maps} (noisy data maps), but for PDF and MF covariance.}%\jl{all cases you meant network size change, augmentation, noise combined? please clarify in caption}}
    \label{fig:cov_PDF_MF_maps}
\end{figure*}

\newtext{For these results we use the following configuration: maps cropped/augmented to $5 \times 5$ sq. degree at $256 \times 256$ pixel (pixel size $ \approx 1.17$ arcmin, with a Gaussian smoothing kernel of $\theta_{\rm s} = 2$ arcmin applied prior to the computation of the $\kappa$-PDF and Minkowski functionals. The power spectrum is computed on the unsmoothed maps}. \newtext{In the augmented-training case, the training set is generated from 3816 $5 \times 5$ sq. degree. After training, $N_{\rm gen} = 4000$maps are drawn from the NF and used to evaluate the summary statistics and covariances in Figures \ref{fig:maps_stats}, \ref{fig:cov_Cell_maps}, and \ref{fig:cov_PDF_MF_maps}.}. Figure~\ref{fig:maps_stats} presents the summary statistics obtained from the NF-generated $\kappa$-maps. For this analysis, we generate $4000$ maps with an angular size of $5\times5~{\rm deg}^2$ and apply a Gaussian smoothing filter with a $2$~arcmin kernel to compute both the PDF and MFs. These results show an improvement in the recovery of the mean and standard deviation of the power spectrum compared to the $256\times256$-pixel map results shown in Figure~\ref{fig:C_ell_maps_resol}. For $C_{\ell}$ even though the variance is slightly overestimated (roughly $10\%$) for large multipole values $l<500$ these allow us to correct for the previously underestimated variance (and covariance) terms at $l>500$. For PDF and MFs results, the results agree with the mean of simulation measurements, however still showing a underestimated variance in some $\nu$ bins. The partial improvement in power spectrum covariance arises from the combination of the studied factors: a larger network architecture with additional multiscale levels, the inclusion of data augmentation in the training set, and the addition of Gaussian noise to the maps during training. Particularly, the addition of noise helps to capture the small-scale variability and yield a more accurate replication of the statistical properties of the weak-lensing convergence field.

The comparison between the replicated and simulated covariance matrices is shown in Figure~\ref{fig:cov_Cell_maps}. We separately assess the effects of increasing the network size, applying data augmentation, and adding pixel noise during training. Overall, both enlarging the network and augmenting the training data improve the recovery of the covariance structure. \newtext{We measure an off-diagonal underestimation of $18\%$ for the larger-network case and $16\%$ for the data-augmentation case. Introducing pixel noise during training is the most effective strategy, reducing the underestimation to the $3\%$ level.} We repeat the covariance analysis for the $\kappa$-PDF and MFs in Figure~\ref{fig:cov_PDF_MF_maps}, \newtext{where a substantial underestimation persists for both non-Gaussian statistics, reaching $13\%$ for the PDF and $30\%$ for the MFs. This poorer performance is expected: the injected pixel noise is Gaussian, and therefore does not contribute to the off-diagonal covariance terms of the PDF or MFs, which are driven by the non-Gaussian structure of the field.} In summary, these results indicate that the accuracy of the replicated summary statistics improves when the network is trained on noisy maps, suggesting that the model is capable of learning the convergence signal when extra stochasticity is added to the training sample.

\begin{figure}
    \centering
    \includegraphics[width=0.95\linewidth]{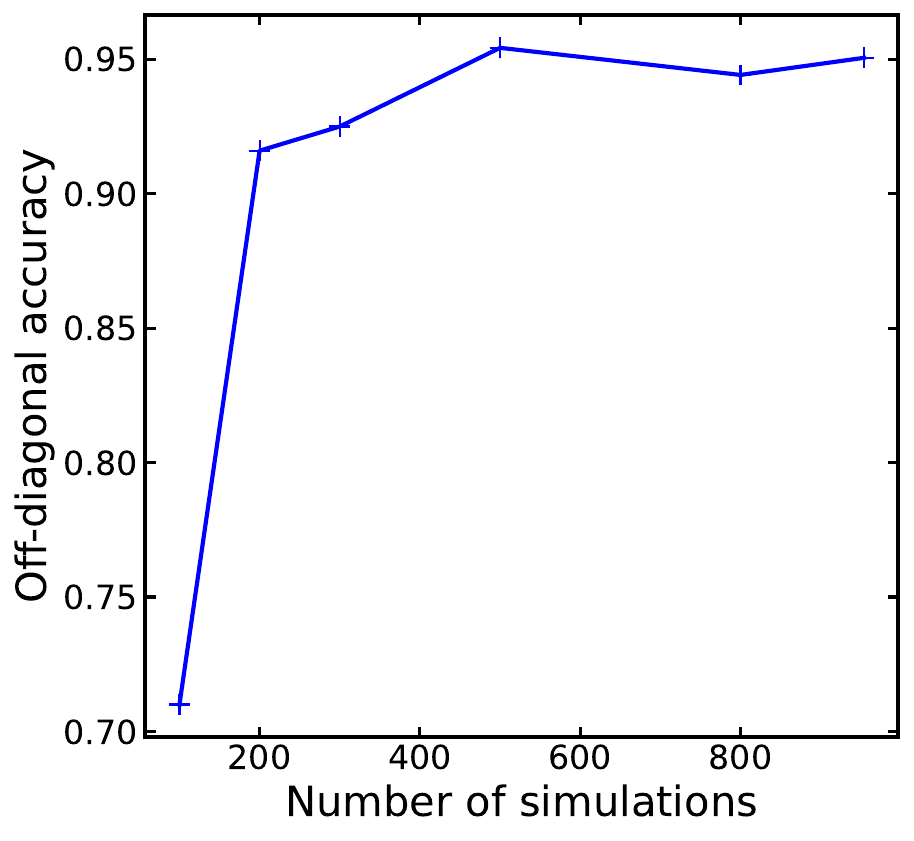}
    \caption{Accuracy of the off-diagonal covariance terms of the convergence power
spectrum $C_{\ell}^{\kappa\kappa}$ measured from normalizing-flow (NF)--generated
maps, as a function of the number of simulations used for training. We construct
six independent training sets containing 100, 200, 300, 500, 800, and 954
simulations, and compare the resulting covariance estimates to that obtained from
the full SLICS simulation suite. The accuracy is computed as
$1 - \left\langle \left| \rho^{\prime} - \rho \right| \right\rangle$, where the
average is taken over the residuals of the off-diagonal correlation coefficients
across the $\ell$ bins used to estimate $C_{\ell}^{\kappa\kappa}$, since these
terms are approximately scale-independent. The absolute value ensures that both
under- and overestimation of the covariance are penalized. The diagonal elements
(variances) are excluded in order to isolate and quantify potential
misestimation of the off-diagonal covariance.}
     %\jl{1. how do you define the off-diag accuracy with just 1 jumber? the average of all the off-diag components? 2. can you separate the 2 effects by comparing off-diag term to the same number of sims? and take out the superficial variance increase, then the curve should increase monotonically? i guess i'm unclear of the denominators used here, if they're the same for all data points} \ja{The number of simulations I change (x-axis) is the how many simulations I use to generate the NF sample. The accuracy is always defined as the covariance (calculated using 954 noisy maps), using either the network and SLICS. Maybe I'm not explaining it well on the text...}
    
    \label{fig:CovAcc}
\end{figure}

Finally, we evaluate the accuracy of the off-diagonal covariance terms as a function of the number of simulations included in the training set, as shown in Figure~\ref{fig:CovAcc}. By varying the number of simulations used to construct the training data, we aim to achieve two objectives: to artificially increase the variance by training on a random subsample of the original dataset, while preserving the mean, since only a few realizations ($\mathcal{O}(100)$) are sufficient to estimate the mean power spectrum accurately. And, to examine how the covariance accuracy behaves for smaller training sets. \newtext{We define the accuracy as $1 - \left< { \left| {\rho^{\prime} - \rho } \right|} \right>$, as the mean of the residual $\rho^{\prime} - \rho$} should be close to zero for the full recovery of the covariance matrix, and possible deviations are either $+1$ or $-1$, then the accuracy is a value between 0 and 1. The results in Figure~\ref{fig:CovAcc} indicate that using approximately \newtext{200} simulations is sufficient to reproduce the SLICS covariance with an average accuracy of about 92\% in the off-diagonal elements, for a $C_{\ell}^{\kappa\kappa}$ like the one in Figure \ref{fig:maps_stats}. The accuracy slightly increases as the number of simulations in the subsample increases, reaching values of \newtext{$\sim95\%$ at 500 simulations. For a larger number of simulations until we approach to the size of the full sample ($\mathcal{O}(1000)$), the accuracy is virtually the same}, as the network learns the full distribution of the dataset and therefore recovers a more representative covariance. Nevertheless, even in this case the accuracy remains close to 95\%, suggesting that the off-diagonal covariance is slightly underestimated even when the network is trained on the complete set of maps. It is useful to note that for small numbers of training simulations \newtext{(close to 100)}, the network is relatively undertrained in comparison to a full sample and has not yet captured the intrinsic variability of the convergence fields. In this regime, the stochasticity in the generated maps arising can introduce an additional scatter that artificially inflates the apparent covariance, \newtext{partially overcompensating for the true estimation of the off-diagonal terms}. As the size of the training set increases, the scatter is reduced as the network predictions become more stable. In the limit of very small training sets, this effect can be understood by considering the network effectively untrained. Only when the network is trained on a sufficiently large fraction of the full simulation suite does it begin to accurately learn the full data distribution, at which point the recovered covariance converges towards its true value.

\section{Conclusions} \label{sec:conclusions}

In this paper, we study the replication of a set of summary statistics of the convergence field using normalizing flow neural networks. To understand how dimensionality affects these measurements, we test the NF generated data at two levels: the summary statistics level and the map level.

For the first level, by using a simple architecture based on the neural spline flow network, the summary statistics are replicated accurately at the percentage level, for the mean and standard deviation (variance). For covariance, the accuracy is close to 5-10\% but depends on the type of summary statistics and the dimensionality of the data vector. We find $C_{\ell}^{\kappa\kappa}$ calculated in up to 50 log-separated $\ell$ multipole bins, and $\kappa$-PDF and MF binned in 20 $\nu$ bins recover the statistics more accurately ($\mathcal{O}(1\%)$). However, we find that once the dimensionality of the data vector to replicate increases, the accuracy of the covariance is largely underestimated, for a fixed network size ($\mathcal{O}(10\%)$).%\jl{last sentence is unclear to me, what do you mean compromised? why does dimensionality matter? how big does the dimensionality needs to be to go wrong? please describe in precise number. e.g. "NSF network reaches percent level accuracy for the mean of all stats, but for the covariance, xx\% accuracy for clkk, xx\% for MF, xx\% for PDF"..}

For the second level, where we train directly on the $\kappa$-maps instead of the summary statistics. We show that for the \newtext{small multiscale network, with one squeeze and split transformation,} %\jl{spell out the full name and network size, as many people only read conclusion}, 
the results lack the correct replication of the $C_{\ell}^{\kappa\kappa}$ variance and covariance by around 25\% and 75\% disagreement. %\jl{"underperform" is ambiguous, you mean underestimated the covariance?}. 
The effect increases when using larger maps and keeping the network fixed, suggesting more multiscale layers are needed. To improve the accuracy performance, we increased the size of the multiscale architecture to incorporate the learning of different scales. \newtext{By increasing the network size, and adding more scale transformations (MS3), the \newtext{discrepancy} of the variance and covariance is reduced at the \newtext{18\%} level.} %\jl{why? do we have to use augmentation? i thought augmentation is a separate thing independent of network size}, 
\newtext{Similar effect is obtained when using data augmentation to increase the size of the training set, recovering the covariance off-diagonal coefficients with \newtext{disagreement} of \newtext{16}\%. The accuracy of the mean, variance and covariance of the summary statistics is largely improved once a noisy data is added to the training set, leading to a \newtext{3}\% \newtext{difference} in the off-diagonal terms for the covariance of the replicated power spectrum. For PDF and MF, NF networks reach $\approx$10-30\% disagreement for covariance, which is still large when compared to data vector replication case.} 

These findings are still consistent with previous applications of machine learning to weak lensing, which demonstrate the potential of generative models—such as GANs, VAEs, and normalizing flows \citep{Borua2024, Mustafa2019,Dai2022,Shirasaki2024,Aoyama2025} to replicate non-Gaussian features of convergence fields and enable fast map generation. However, we emphasize how the limited information used for training can compromise the accuracy of the off-diagonal terms of the covariance for different summary statistics, indicating that further steps must be taken before using generated maps for cosmological inference, as underestimated covariance can lead to overconfident contours. A simple explanation for this behavior is the poor scaling of NF networks when a limited sample of simulations is used when train high-dimensional maps. This suggests that the sub-arcmin Requirements of stage-IV weak lensing surveys require large models, with an effective understanding of noise and data augmentation techniques, as shown in different studies \cite{Shirasaki2021,zhong2024}. In particular, diffusion model architectures~\cite{Boruah2025}, where the $\kappa$ signal and pixel noise must be disentangled, can benefit from these data augmentation techniques.\newtext{The relevance of our findings lies on the diagnostics of NF-generated maps recovering the complete underlying distribution of the data. Generative models for map statistics are increasingly used as fast emulators, as training sets for inference pipelines, and as tools for exploring rare events or augmenting limited simulation suites. For all of these uses a persistent failure mode that can lead to overconfident posteriors if the NF outputs are used naively. We encourage the use of data augmentation and noise injection as mitigation techniques to overcome these issues.}

%\newtext{Add paragraph discussing other studies (convergence GAN paper, DM paper, NF papers) and how this work can help to improve them...}.

\begin{acknowledgments}
JA acknowledges useful discussions with Ben Horowitz and Linda Blot. We thank Joachim Harnois-Déraps for providing the SLICS simulations. JA is supported by JSPS KAKENHI Grant JP23K19064. LT is supported by JSPS KAKENHI Grant 24K22878. JL is supported by JSPS KAKENHI Grants 23H00107, 25H00403. We acknowledge the Texas Advanced Computing Center (TACC) at The University of Texas at Austin for providing grid resources that have contributed to our work, including the Anvil machine for GPU computation. This research used computing resources at Kavli IPMU. The Kavli IPMU is supported by the WPI (World Premier International Research Center) Initiative of the MEXT (Japanese Ministry of Education, Culture, Sports, Science and Technology).
 
\end{acknowledgments}

%\appendix

%\section{Multiscale transformations for weak lensing images} \label{app:MS}

%We detail in this appendix the normalizing flow multiscale network and the data augmentation techniques used for the training.

%\section{Data augmentation} \label{app:DA}

%\nocite{*}

\bibliography{apssamp}% Produces the bibliography via BibTeX.

@ARTICLE{Zuntz2021,
       author = {{Zuntz}, Joe and {Lanusse}, Fran{\c{c}}ois and {Malz}, Alex I. and {Wright}, Angus H. and {Slosar}, An{\v{z}}e and {Abolfathi}, Bela and {Alonso}, David and {Bault}, Abby and {Bom}, Cl{\'e}cio R. and {Brescia}, Massimo and {Broussard}, Adam and {Campagne}, Jean-Eric and {Cavuoti}, Stefano and {Cypriano}, Eduardo S. and {Fraga}, Bernardo M.~O. and {Gawiser}, Eric and {Gonzalez}, Elizabeth J. and {Green}, Dylan and {Hatfield}, Peter and {Iyer}, Kartheik and {Kirkby}, David and {Nicola}, Andrina and {Nourbakhsh}, Erfan and {Park}, Andy and {Teixeira}, Gabriel and {Heitmann}, Katrin and {Kovacs}, Eve and {Mao}, Yao-Yuan and {LSST Dark Energy Science Collaboration}},
        title = "{The LSST-DESC 3x2pt Tomography Optimization Challenge}",
      journal = {The Open Journal of Astrophysics},
         year = 2021,
        month = oct,
       volume = {4},
       number = {1},
          eid = {13},
        pages = {13},
          doi = {10.21105/astro.2108.13418},
archivePrefix = {arXiv},
       eprint = {2108.13418},
 primaryClass = {astro-ph.IM},
       adsurl = {https://ui.adsabs.harvard.edu/abs/2021OJAp....4E..13Z}
}

@article{PearsonSamushia2016,
    author = {Pearson, David W. and Samushia, Lado},
    title = {Estimating the power spectrum covariance matrix with fewer mock samples},
    journal = {Monthly Notices of the Royal Astronomical Society},
    volume = {457},
    number = {1},
    pages = {993-999},
    year = {2016},
    month = {01},
    issn = {0035-8711},
    doi = {10.1093/mnras/stw062},
    url = {https://doi.org/10.1093/mnras/stw062},
    eprint = {https://academic.oup.com/mnras/article-pdf/457/1/993/18169038/stw062.pdf},
}

@ARTICLE{Pope2008,
       author = {{Pope}, Adrian C. and {Szapudi}, Istv{\'a}n},
        title = "{Shrinkage estimation of the power spectrum covariance matrix}",
      journal = {\mnras},
         year = 2008,
        month = sep,
       volume = {389},
       number = {2},
        pages = {766-774},
          doi = {10.1111/j.1365-2966.2008.13561.x},
archivePrefix = {arXiv},
       eprint = {0711.2509},
 primaryClass = {astro-ph},
       adsurl = {https://ui.adsabs.harvard.edu/abs/2008MNRAS.389..766P}
}

@ARTICLE{Heymans2021,
       author = {{Heymans}, Catherine and {Tr{\"o}ster}, Tilman and {Asgari}, Marika and {Blake}, Chris and {Hildebrandt}, Hendrik and {Joachimi}, Benjamin and {Kuijken}, Konrad and {Lin}, Chieh-An and {S{\'a}nchez}, Ariel G. and {van den Busch}, Jan Luca and {Wright}, Angus H. and {Amon}, Alexandra and {Bilicki}, Maciej and {de Jong}, Jelte and {Crocce}, Martin and {Dvornik}, Andrej and {Erben}, Thomas and {Fortuna}, Maria Cristina and {Getman}, Fedor and {Giblin}, Benjamin and {Glazebrook}, Karl and {Hoekstra}, Henk and {Joudaki}, Shahab and {Kannawadi}, Arun and {K{\"o}hlinger}, Fabian and {Lidman}, Chris and {Miller}, Lance and {Napolitano}, Nicola R. and {Parkinson}, David and {Schneider}, Peter and {Shan}, HuanYuan and {Valentijn}, Edwin A. and {Verdoes Kleijn}, Gijs and {Wolf}, Christian},
        title = "{KiDS-1000 Cosmology: Multi-probe weak gravitational lensing and spectroscopic galaxy clustering constraints}",
      journal = {\aap},
         year = 2021,
        month = feb,
       volume = {646},
          eid = {A140},
        pages = {A140},
          doi = {10.1051/0004-6361/202039063},
archivePrefix = {arXiv},
       eprint = {2007.15632},
 primaryClass = {astro-ph.CO},
       adsurl = {https://ui.adsabs.harvard.edu/abs/2021A&A...646A.140H}
}

@ARTICLE{Papamakarios2017,
       author = {{Papamakarios}, George and {Pavlakou}, Theo and {Murray}, Iain},
        title = "{Masked Autoregressive Flow for Density Estimation}",
      journal = {arXiv e-prints},
         year = 2017,
        month = may,
          eid = {arXiv:1705.07057},
        pages = {arXiv:1705.07057},
          doi = {10.48550/arXiv.1705.07057},
archivePrefix = {arXiv},
       eprint = {1705.07057},
 primaryClass = {stat.ML},
       adsurl = {https://ui.adsabs.harvard.edu/abs/2017arXiv170507057P}
}

@ARTICLE{Uhlemann2020,
       author = {{Uhlemann}, Cora and {Friedrich}, Oliver and {Villaescusa-Navarro}, Francisco and {Banerjee}, Arka and {Codis}, Sandrine},
        title = "{Fisher for complements: extracting cosmology and neutrino mass from the counts-in-cells PDF}",
      journal = {\mnras},
         year = 2020,
        month = jul,
       volume = {495},
       number = {4},
        pages = {4006-4027},
          doi = {10.1093/mnras/staa1155},
archivePrefix = {arXiv},
       eprint = {1911.11158},
 primaryClass = {astro-ph.CO},
       adsurl = {https://ui.adsabs.harvard.edu/abs/2020MNRAS.495.4006U}
}

@article{Bartelmann2001,
  title={Weak gravitational lensing},
  author={Bartelmann, M. and Schneider, P.},
  journal={Physics Reports},
  volume={340},
  number={4-5},
  pages={291--472},
  year={2001}
}

@article{Kilbinger2015,
  title={Cosmology with cosmic shear observations: a review},
  author={Kilbinger, M.},
  journal={Reports on Progress in Physics},
  volume={78},
  number={8},
  pages={086901},
  year={2015}
}

@article{Huterer2010,
  title={Weak lensing and dark energy},
  author={Huterer, D.},
  journal={General Relativity and Gravitation},
  volume={42},
  pages={2177–2195},
  year={2010}
}

@article{DES2022,
  title = {Dark Energy Survey Year 3 results: Cosmology from cosmic shear and robustness to data calibration},
  author = {Amon, A. and Gruen, D. and Troxel, M. A. and MacCrann, N. and et al.},
  collaboration = {DES Collaboration},
  journal = {Phys. Rev. D},
  volume = {105},
  issue = {2},
  pages = {023514},
  numpages = {43},
  year = {2022},
  month = {Jan},
  publisher = {American Physical Society},
  doi = {10.1103/PhysRevD.105.023514},
  url = {https://link.aps.org/doi/10.1103/PhysRevD.105.023514}
}

@article{Hikage2019,
  title={Cosmology from cosmic shear power spectra with Subaru Hyper Suprime-Cam first-year data},
  author={Hikage, C. et al.},
  journal={Publications of the Astronomical Society of Japan},
  volume={71},
  issue={2},
  pages={43},
  year={2019}
}

@techreport{Laureijs2011,
  title={Euclid Definition Study Report},
  author={Laureijs, R. et al.},
  institution={ESA/SRE},
  number={2011-12},
  year={2011}
}

@ARTICLE{LSST2009,
       author = {{LSST Science Collaboration} },
        title = "{LSST Science Book, Version 2.0}",
      journal = {arXiv e-prints},
         year = 2009,
        month = dec,
          eid = {arXiv:0912.0201},
        pages = {arXiv:0912.0201},
          doi = {10.48550/arXiv.0912.0201},
archivePrefix = {arXiv},
       eprint = {0912.0201},
 primaryClass = {astro-ph.IM},
       adsurl = {https://ui.adsabs.harvard.edu/abs/2009arXiv0912.0201L}
}

@article{KaiserSquires1993,
  title={Mapping the dark matter with weak gravitational lensing},
  author={Kaiser, N. and Squires, G.},
  journal={The Astrophysical Journal},
  volume={404},
  pages={441--450},
  year={1993}
}

@article{Takahashi2017,
  title={Full-sky gravitational lensing simulation for large-area galaxy surveys and cosmic microwave background experiments},
  author={Takahashi, R. et al.},
  journal={The Astrophysical Journal},
  volume={850},
  number={1},
  pages={24},
  year={2017}
}

@article{Jeffrey2021,
  title={Mass-mapping meets machine learning: removing shape noise with generative models},
  author={Jeffrey, N. et al.},
  journal={Monthly Notices of the Royal Astronomical Society},
  volume={505},
  number={3},
  pages={4626--4645},
  year={2021}
}

@article{Ravanbakhsh2017,
       author = {{Ravanbakhsh}, Siamak and {Oliva}, Junier and {Fromenteau}, Sebastien and {Price}, Layne C. and {Ho}, Shirley and {Schneider}, Jeff and {Poczos}, Barnabas},
        title = "{Estimating Cosmological Parameters from the Dark Matter Distribution}",
      journal = {arXiv e-prints},
         year = 2017,
        month = nov,
          eid = {arXiv:1711.02033},
        pages = {arXiv:1711.02033},
          doi = {10.48550/arXiv.1711.02033},
archivePrefix = {arXiv},
       eprint = {1711.02033},
 primaryClass = {astro-ph.CO},
       adsurl = {https://ui.adsabs.harvard.edu/abs/2017arXiv171102033R}
}

@article{Remy2020,
  title={CAMELS: Cosmology and Astrophysics with Machine-learning Simulations},
  author={Villaescusa-Navarro, F. et al.},
  journal={The Astrophysical Journal},
  volume={915},
  number={1},
  pages={71},
  year={2021}
}

@article{mandelbaum2018weak,
  title={Weak lensing for precision cosmology},
  author={Mandelbaum, Rachel},
  journal={Annual Review of Astronomy and Astrophysics},
  volume={56},
  pages={393--433},
  year={2018}
}

@article{DES2021cosmicshear,
  title={Dark Energy Survey Year 3 results: Cosmology from cosmic shear and robustness to modeling uncertainty},
  author={Amon, Alexandra and others},
  journal={Physical Review D},
  volume={105},
  number={2},
  pages={023520},
  year={2022}
}

@article{asgari2021kids,
  title={KiDS-1000 Cosmology: Cosmic shear constraints and comparison between two point statistics},
  author={Asgari, Marika and others},
  journal={Astronomy \& Astrophysics},
  volume={645},
  pages={A104},
  year={2021}
}

@article{hikage2019hsc,
  title={Cosmology from cosmic shear power spectra with Subaru Hyper Suprime-Cam first-year data},
  author={Hikage, Chiaki and others},
  journal={Publications of the Astronomical Society of Japan},
  volume={71},
  number={2},
  pages={43},
  year={2019}
}

@article{Armijo2025,
    author = {Armijo, Joaquin and Marques, Gabriela A and Novaes, Camila P and Thiele, Leander and Cowell, Jessica A and Grandón, Daniela and Shirasaki, Masato and Liu, Jia},
    title = {Cosmological constraints using Minkowski functionals from the first year data of the Hyper Suprime-Cam},
    journal = {Monthly Notices of the Royal Astronomical Society},
    pages = {staf257},
    year = {2025},
    month = {02},
    issn = {0035-8711},
    doi = {10.1093/mnras/staf257},
    url = {https://doi.org/10.1093/mnras/staf257},
    eprint = {https://academic.oup.com/mnras/advance-article-pdf/doi/10.1093/mnras/staf257/61870752/staf257.pdf},
}

@article{harnois2018slics,
  title={SLICS: A suite of cosmological N-body simulations for lensing covariance estimation and beyond},
  author={Harnois-Déraps, Joachim and others},
  journal={Monthly Notices of the Royal Astronomical Society},
  volume={481},
  number={1},
  pages={1337--1350},
  year={2018}
}

@article{petri2016cosmology,
  title={Cosmology with the convergence probability distribution function and moments in weak lensing surveys},
  author={Petri, Andrea},
  journal={Physical Review D},
  volume={94},
  number={6},
  pages={063534},
  year={2016}
}

@article{kratochvil2012minkowski,
  title={Probing cosmology with weak lensing Minkowski functionals},
  author={Kratochvil, Jan M. and Lim, Eugene A. and Wang, Shuang},
  journal={Physical Review D},
  volume={85},
  number={10},
  pages={103513},
  year={2012}
}

@article{munshi2012minkowski,
  title={Minkowski Functionals in Cosmology},
  author={Munshi, Dipak and others},
  journal={Monthly Notices of the Royal Astronomical Society},
  volume={419},
  number={1},
  pages={536--555},
  year={2012}
}

@ARTICLE{Dinh2016,
       author = {{Dinh}, Laurent and {Sohl-Dickstein}, Jascha and {Bengio}, Samy},
        title = "{Density estimation using Real NVP}",
      journal = {arXiv e-prints},
         year = 2016,
        month = may,
          eid = {arXiv:1605.08803},
        pages = {arXiv:1605.08803},
          doi = {10.48550/arXiv.1605.08803},
archivePrefix = {arXiv},
       eprint = {1605.08803},
 primaryClass = {cs.LG},
       adsurl = {https://ui.adsabs.harvard.edu/abs/2016arXiv160508803D}
}

@INPROCEEDINGS{Dai_Seljak2023,
       author = {{Dai}, Biwei and {Seljak}, Uros},
        title = "{Multiscale Flow for Robust and Optimal Cosmological Analysis}",
    booktitle = {Machine Learning for Astrophysics},
         year = 2023,
        month = jul,
          eid = {10},
        pages = {10},
          doi = {10.48550/arXiv.2306.04689},
archivePrefix = {arXiv},
       eprint = {2306.04689},
 primaryClass = {astro-ph.CO},
       adsurl = {https://ui.adsabs.harvard.edu/abs/2023mla..confE..10D}
}

@article{Fluri2019,
  title={Cosmological constraints from noisy convergence maps through deep learning},
  author={Fluri, Joris and Kacprzak, Tomasz and Lucchi, Aurelien and Amara, Adam and others},
  journal={Physical Review D},
  volume={100},
  number={6},
  pages={063514},
  year={2019},
  doi={10.1103/PhysRevD.100.063514}
}

@ARTICLE{Munshi2000,
       author = {{Munshi}, Dipak and {Jain}, Bhuvnesh},
        title = "{The statistics of weak lensing at small angular scales: probability distribution function}",
      journal = {\mnras},
         year = 2000,
        month = oct,
       volume = {318},
       number = {1},
        pages = {109-123},
          doi = {10.1046/j.1365-8711.2000.03694.x},
archivePrefix = {arXiv},
       eprint = {astro-ph/9911502},
 primaryClass = {astro-ph},
       adsurl = {https://ui.adsabs.harvard.edu/abs/2000MNRAS.318..109M}
}

@ARTICLE{Huterer2018,
       author = {{Huterer}, Dragan and {Shafer}, Daniel L.},
        title = "{Dark energy two decades after: observables, probes, consistency tests}",
      journal = {Reports on Progress in Physics},
         year = 2018,
        month = jan,
       volume = {81},
       number = {1},
          eid = {016901},
        pages = {016901},
          doi = {10.1088/1361-6633/aa997e},
archivePrefix = {arXiv},
       eprint = {1709.01091},
 primaryClass = {astro-ph.CO},
       adsurl = {https://ui.adsabs.harvard.edu/abs/2018RPPh...81a6901H}
}

@ARTICLE{Pires2009,
       author = {{Pires}, S. and {Starck}, J. -L. and {Amara}, A. and {Teyssier}, R. and {R{\'e}fr{\'e}gier}, A. and {Fadili}, J.},
        title = "{FAst STatistics for weak Lensing (FASTLens): fast method for weak lensing statistics and map making}",
      journal = {\mnras},
         year = 2009,
        month = may,
       volume = {395},
       number = {3},
        pages = {1265-1279},
          doi = {10.1111/j.1365-2966.2009.14625.x},
archivePrefix = {arXiv},
       eprint = {0804.4068},
 primaryClass = {astro-ph},
       adsurl = {https://ui.adsabs.harvard.edu/abs/2009MNRAS.395.1265P}
}

@ARTICLE{ValeWhite2003,
       author = {{Vale}, Chris and {White}, Martin},
        title = "{Simulating Weak Lensing by Large-Scale Structure}",
      journal = {\apj},
         year = 2003,
        month = aug,
       volume = {592},
       number = {2},
        pages = {699-709},
          doi = {10.1086/375867},
archivePrefix = {arXiv},
       eprint = {astro-ph/0303555},
 primaryClass = {astro-ph},
       adsurl = {https://ui.adsabs.harvard.edu/abs/2003ApJ...592..699V}
}

@ARTICLE{Petri2016,
       author = {{Petri}, Andrea and {Haiman}, Zolt{\'a}n and {May}, Morgan},
        title = "{Sample variance in weak lensing: How many simulations are required?}",
      journal = {\prd},
         year = 2016,
        month = mar,
       volume = {93},
       number = {6},
          eid = {063524},
        pages = {063524},
          doi = {10.1103/PhysRevD.93.063524},
archivePrefix = {arXiv},
       eprint = {1601.06792},
 primaryClass = {astro-ph.CO},
       adsurl = {https://ui.adsabs.harvard.edu/abs/2016PhRvD..93f3524P}
}

@ARTICLE{Castiblanco2024,
       author = {{Castiblanco}, Lina and {Uhlemann}, Cora and {Harnois-D{\'e}raps}, Joachim and {Barthelemy}, Alexandre},
        title = "{Unleashing cosmic shear information with the tomographic weak lensing PDF}",
      journal = {The Open Journal of Astrophysics},
         year = 2024,
        month = jul,
       volume = {7},
          eid = {59},
        pages = {59},
          doi = {10.33232/001c.121302},
archivePrefix = {arXiv},
       eprint = {2405.09651},
 primaryClass = {astro-ph.CO},
       adsurl = {https://ui.adsabs.harvard.edu/abs/2024OJAp....7E..59C}
}

@ARTICLE{Thiele2023,
       author = {{Thiele}, Leander and {Marques}, Gabriela A. and {Liu}, Jia and {Shirasaki}, Masato},
        title = "{Cosmological constraints from the Subaru Hyper Suprime-Cam year 1 shear catalogue lensing convergence probability distribution function}",
      journal = {\prd},
         year = 2023,
        month = dec,
       volume = {108},
       number = {12},
          eid = {123526},
        pages = {123526},
          doi = {10.1103/PhysRevD.108.123526},
       adsurl = {https://ui.adsabs.harvard.edu/abs/2023PhRvD.108l3526T}
}

@ARTICLE{Marques_2024,
       author = {{Marques}, Gabriela A. and {Liu}, Jia and {Shirasaki}, Masato and {Thiele}, Leander and {Grand{\'o}n}, Daniela and {Huffenberger}, Kevin M. and {Cheng}, Sihao and {Harnois-D{\'e}raps}, Joachim and {Osato}, Ken and {Coulton}, William R.},
        title = "{Cosmology from weak lensing peaks and minima with Subaru Hyper Suprime-Cam Survey first-year data}",
      journal = {\mnras},
         year = 2024,
        month = mar,
       volume = {528},
       number = {3},
        pages = {4513-4527},
          doi = {10.1093/mnras/stae098},
archivePrefix = {arXiv},
       eprint = {2308.10866},
 primaryClass = {astro-ph.CO},
       adsurl = {https://ui.adsabs.harvard.edu/abs/2024MNRAS.528.4513M}
}

@ARTICLE{Cheng_2024,
       author = {{Cheng}, Sihao and {Marques}, Gabriela A. and {Grand{\'o}n}, Daniela and {Thiele}, Leander and {Shirasaki}, Masato and {M{\'e}nard}, Brice and {Liu}, Jia},
        title = "{Cosmological constraints from weak lensing scattering transform using HSC Y1 data}",
      journal = {arXiv e-prints},
         year = 2024,
        month = apr,
          eid = {arXiv:2404.16085},
        pages = {arXiv:2404.16085},
          doi = {10.48550/arXiv.2404.16085},
archivePrefix = {arXiv},
       eprint = {2404.16085},
 primaryClass = {astro-ph.CO},
       adsurl = {https://ui.adsabs.harvard.edu/abs/2024arXiv240416085C}
}

@ARTICLE{Novaes2024,
       author = {{Novaes}, Camila P. and {Thiele}, Leander and {Armijo}, Joaquin and {Cheng}, Sihao and {Cowell}, Jessica A. and {Marques}, Gabriela A. and {Ferreira}, Elisa G.~M. and {Shirasaki}, Masato and {Osato}, Ken and {Liu}, Jia},
        title = "{Cosmology from HSC Y1 weak lensing data with combined higher-order statistics and simulation-based inference}",
      journal = {\prd},
         year = 2025,
        month = apr,
       volume = {111},
       number = {8},
          eid = {083510},
        pages = {083510},
          doi = {10.1103/PhysRevD.111.083510},
archivePrefix = {arXiv},
       eprint = {2409.01301},
 primaryClass = {astro-ph.CO},
       adsurl = {https://ui.adsabs.harvard.edu/abs/2025PhRvD.111h3510N}
}

@ARTICLE{Durkan2019,
       author = {{Durkan}, Conor and {Bekasov}, Artur and {Murray}, Iain and {Papamakarios}, George},
        title = "{Neural Spline Flows}",
      journal = {arXiv e-prints},
         year = 2019,
        month = jun,
          eid = {arXiv:1906.04032},
        pages = {arXiv:1906.04032},
          doi = {10.48550/arXiv.1906.04032},
archivePrefix = {arXiv},
       eprint = {1906.04032},
 primaryClass = {stat.ML},
       adsurl = {https://ui.adsabs.harvard.edu/abs/2019arXiv190604032D}
}

@ARTICLE{Kingma2018,
       author = {{Kingma}, Diederik P. and {Dhariwal}, Prafulla},
        title = "{Glow: Generative Flow with Invertible 1x1 Convolutions}",
      journal = {arXiv e-prints},
         year = 2018,
        month = jul,
          eid = {arXiv:1807.03039},
        pages = {arXiv:1807.03039},
          doi = {10.48550/arXiv.1807.03039},
archivePrefix = {arXiv},
       eprint = {1807.03039},
 primaryClass = {stat.ML},
       adsurl = {https://ui.adsabs.harvard.edu/abs/2018arXiv180703039K}
}

@ARTICLE{Boruah2025,
       author = {{Boruah}, Supranta S. and {Jacob}, Michael and {Jain}, Bhuvnesh},
        title = "{Diffusion-based mass map reconstruction from weak lensing data}",
      journal = {\prd},
         year = 2025,
        month = apr,
       volume = {111},
       number = {8},
          eid = {083542},
        pages = {083542},
          doi = {10.1103/PhysRevD.111.083542},
archivePrefix = {arXiv},
       eprint = {2502.04158},
 primaryClass = {astro-ph.CO},
       adsurl = {https://ui.adsabs.harvard.edu/abs/2025PhRvD.111h3542B}
}

@ARTICLE{Mustafa2019,
       author = {{Mustafa}, Mustafa and {Bard}, Deborah and {Bhimji}, Wahid and {Luki{\'c}}, Zarija and {Al-Rfou}, Rami and {Kratochvil}, Jan M.},
        title = "{CosmoGAN: creating high-fidelity weak lensing convergence maps using Generative Adversarial Networks}",
      journal = {Computational Astrophysics and Cosmology},
         year = 2019,
        month = may,
       volume = {6},
       number = {1},
          eid = {1},
        pages = {1},
          doi = {10.1186/s40668-019-0029-9},
archivePrefix = {arXiv},
       eprint = {1706.02390},
 primaryClass = {astro-ph.IM},
       adsurl = {https://ui.adsabs.harvard.edu/abs/2019ComAC...6....1M}
}

@ARTICLE{Massey2013,
       author = {{Massey}, Richard and {Hoekstra}, Henk and {Kitching}, Thomas and {Rhodes}, Jason and {Cropper}, Mark and {Amiaux}, J{\'e}r{\^o}me and {Harvey}, David and {Mellier}, Yannick and {Meneghetti}, Massimo and {Miller}, Lance and {Paulin-Henriksson}, St{\'e}phane and {Pires}, Sandrine and {Scaramella}, Roberto and {Schrabback}, Tim},
        title = "{Origins of weak lensing systematics, and requirements on future instrumentation (or knowledge of instrumentation)}",
      journal = {\mnras},
         year = 2013,
        month = feb,
       volume = {429},
       number = {1},
        pages = {661-678},
          doi = {10.1093/mnras/sts371},
archivePrefix = {arXiv},
       eprint = {1210.7690},
 primaryClass = {astro-ph.CO},
       adsurl = {https://ui.adsabs.harvard.edu/abs/2013MNRAS.429..661M}
}

@ARTICLE{Dai2022,
       author = {{Dai}, Biwei and {Seljak}, Uro{\v{s}}},
        title = "{Translation and rotation equivariant normalizing flow (TRENF) for optimal cosmological analysis}",
      journal = {\mnras},
         year = 2022,
        month = oct,
       volume = {516},
       number = {2},
        pages = {2363-2373},
          doi = {10.1093/mnras/stac2010},
archivePrefix = {arXiv},
       eprint = {2202.05282},
 primaryClass = {astro-ph.CO},
       adsurl = {https://ui.adsabs.harvard.edu/abs/2022MNRAS.516.2363D}
}

@ARTICLE{Gatti2022,
       author = {{Gatti}, M. and {Jain}, B. and {Chang}, C. and {Raveri}, M. and {Z{\"u}rcher}, D. and {Secco}, L. et al. {DES Collaboration}},
        title = "{Dark Energy Survey Year 3 results: Cosmology with moments of weak lensing mass maps}",
      journal = {\prd},
         year = 2022,
        month = oct,
       volume = {106},
       number = {8},
          eid = {083509},
        pages = {083509},
          doi = {10.1103/PhysRevD.106.083509},
archivePrefix = {arXiv},
       eprint = {2110.10141},
 primaryClass = {astro-ph.CO},
       adsurl = {https://ui.adsabs.harvard.edu/abs/2022PhRvD.106h3509G}
}

@ARTICLE{zhong2024,
       author = {{Zhong}, Kunhao and {Gatti}, Marco and {Jain}, Bhuvnesh},
        title = "{Improving convolutional neural networks for cosmological fields with random permutation}",
      journal = {\prd},
         year = 2024,
        month = aug,
       volume = {110},
       number = {4},
          eid = {043535},
        pages = {043535},
          doi = {10.1103/PhysRevD.110.043535},
archivePrefix = {arXiv},
       eprint = {2403.01368},
 primaryClass = {astro-ph.CO},
       adsurl = {https://ui.adsabs.harvard.edu/abs/2024PhRvD.110d3535Z}
}

@ARTICLE{Ribli2019,
       author = {{Ribli}, Dezs{\H{o}} and {Pataki}, B{\'a}lint {\'A}rmin and {Zorrilla Matilla}, Jos{\'e} Manuel and {Hsu}, Daniel and {Haiman}, Zolt{\'a}n and {Csabai}, Istv{\'a}n},
        title = "{Weak lensing cosmology with convolutional neural networks on noisy data}",
      journal = {\mnras},
         year = 2019,
        month = dec,
       volume = {490},
       number = {2},
        pages = {1843-1860},
          doi = {10.1093/mnras/stz2610},
archivePrefix = {arXiv},
       eprint = {1902.03663},
 primaryClass = {astro-ph.CO},
       adsurl = {https://ui.adsabs.harvard.edu/abs/2019MNRAS.490.1843R}
}

@ARTICLE{Whitney2024,
       author = {{Whitney}, Jessica and {Liaudat}, Tob{\'\i}as and {Price}, Matt and {Mars}, Matthijs and {McEwen}, Jason D.},
        title = "{Using conditional GANs for convergence map reconstruction with uncertainties}",
      journal = {arXiv e-prints},
         year = 2024,
        month = may,
          eid = {arXiv:2406.15424},
        pages = {arXiv:2406.15424},
          doi = {10.48550/arXiv.2406.15424},
archivePrefix = {arXiv},
       eprint = {2406.15424},
 primaryClass = {astro-ph.CO},
       adsurl = {https://ui.adsabs.harvard.edu/abs/2024arXiv240615424W}
}

@ARTICLE{Whitney2024b,
       author = {{Whitney}, Jessica J. and {Liaudat}, Tob{\'\i}as I. and {Price}, Matthew A. and {Mars}, Matthijs and {McEwen}, Jason D.},
        title = "{Generative modelling for mass-mapping with fast uncertainty quantification}",
      journal = {arXiv e-prints},
         year = 2024,
        month = oct,
          eid = {arXiv:2410.24197},
        pages = {arXiv:2410.24197},
          doi = {10.48550/arXiv.2410.24197},
archivePrefix = {arXiv},
       eprint = {2410.24197},
 primaryClass = {astro-ph.CO},
       adsurl = {https://ui.adsabs.harvard.edu/abs/2024arXiv241024197W}
}

@ARTICLE{Borua2024,
       author = {{Boruah}, Supranta S. and {Fiedorowicz}, Pier and {Garcia}, Rafael and {Coulton}, William R. and {Rozo}, Eduardo and {Fabbian}, Giulio},
        title = "{GANSky -- fast curved sky weak lensing simulations using Generative Adversarial Networks}",
      journal = {arXiv e-prints},
         year = 2024,
        month = jun,
          eid = {arXiv:2406.05867},
        pages = {arXiv:2406.05867},
          doi = {10.48550/arXiv.2406.05867},
archivePrefix = {arXiv},
       eprint = {2406.05867},
 primaryClass = {astro-ph.CO},
       adsurl = {https://ui.adsabs.harvard.edu/abs/2024arXiv240605867B}
}

@ARTICLE{Shirasaki2024,
       author = {{Shirasaki}, Masato and {Ikeda}, Shiro},
        title = "{Neural style transfer of weak lensing mass maps}",
      journal = {The Open Journal of Astrophysics},
         year = 2024,
        month = may,
       volume = {7},
          eid = {42},
        pages = {42},
          doi = {10.33232/001c.118104},
archivePrefix = {arXiv},
       eprint = {2310.17141},
 primaryClass = {astro-ph.CO},
       adsurl = {https://ui.adsabs.harvard.edu/abs/2024OJAp....7E..42S}
}

@ARTICLE{Shirasaki2021,
       author = {{Shirasaki}, Masato and {Moriwaki}, Kana and {Oogi}, Taira and {Yoshida}, Naoki and {Ikeda}, Shiro and {Nishimichi}, Takahiro},
        title = "{Noise reduction for weak lensing mass mapping: an application of generative adversarial networks to Subaru Hyper Suprime-Cam first-year data}",
      journal = {\mnras},
         year = 2021,
        month = jun,
       volume = {504},
       number = {2},
        pages = {1825-1839},
          doi = {10.1093/mnras/stab982},
archivePrefix = {arXiv},
       eprint = {1911.12890},
 primaryClass = {astro-ph.CO},
       adsurl = {https://ui.adsabs.harvard.edu/abs/2021MNRAS.504.1825S}
}

@article{Shirasaki2019,
  title = {Denoising weak lensing mass maps with deep learning},
  author = {Shirasaki, Masato and Yoshida, Naoki and Ikeda, Shiro},
  journal = {Phys. Rev. D},
  volume = {100},
  issue = {4},
  pages = {043527},
  numpages = {14},
  year = {2019},
  month = {Aug},
  publisher = {American Physical Society},
  doi = {10.1103/PhysRevD.100.043527},
  url = {https://link.aps.org/doi/10.1103/PhysRevD.100.043527}
}

@article{Gupta2018,
  title = {Non-Gaussian information from weak lensing data via deep learning},
  author = {Gupta, Arushi and Matilla, Jos\'e Manuel Zorrilla and Hsu, Daniel and Haiman, Zolt\'an},
  journal = {Phys. Rev. D},
  volume = {97},
  issue = {10},
  pages = {103515},
  numpages = {15},
  year = {2018},
  month = {May},
  publisher = {American Physical Society},
  doi = {10.1103/PhysRevD.97.103515},
  url = {https://link.aps.org/doi/10.1103/PhysRevD.97.103515}
}

@article{Aoyama2025,
    author = {Aoyama, Shohei D and Osato, Ken and Shirasaki, Masato},
    title = {Denoising weak lensing mass maps with the diffusion model: Systematic comparison with the generative adversarial network},
    journal = {Publications of the Astronomical Society of Japan},
    volume = {78},
    number = {3},
    pages = {1163-1180},
    year = {2026},
    month = {06},
    issn = {2053-051X},
    doi = {10.1093/pasj/psag052},
    url = {https://doi.org/10.1093/pasj/psag052},
    eprint = {https://academic.oup.com/pasj/article-pdf/78/3/1163/68271439/psag052.pdf},
}

@ARTICLE{Perraudin2020,
       author = {{Perraudin}, Nathana{\"e}l and {Marcon}, Sandro and {Lucchi}, Aurelien and {Kacprzak}, Tomasz},
        title = "{Emulation of cosmological mass maps with conditional generative adversarial networks}",
      journal = {arXiv e-prints},
         year = 2020,
        month = apr,
          eid = {arXiv:2004.08139},
        pages = {arXiv:2004.08139},
          doi = {10.48550/arXiv.2004.08139},
archivePrefix = {arXiv},
       eprint = {2004.08139},
 primaryClass = {astro-ph.CO},
       adsurl = {https://ui.adsabs.harvard.edu/abs/2020arXiv200408139P}
}

@ARTICLE{Rodriguez2018,
       author = {{Rodr{\'\i}guez}, Andres C. and {Kacprzak}, Tomasz and {Lucchi}, Aurelien and {Amara}, Adam and {Sgier}, Rapha{\"e}l and {Fluri}, Janis and {Hofmann}, Thomas and {R{\'e}fr{\'e}gier}, Alexandre},
        title = "{Fast cosmic web simulations with generative adversarial networks}",
      journal = {Computational Astrophysics and Cosmology},
         year = 2018,
        month = nov,
       volume = {5},
       number = {1},
          eid = {4},
        pages = {4},
          doi = {10.1186/s40668-018-0026-4},
archivePrefix = {arXiv},
       eprint = {1801.09070},
 primaryClass = {astro-ph.CO},
       adsurl = {https://ui.adsabs.harvard.edu/abs/2018ComAC...5....4R}
}

@ARTICLE{Troster2019,
       author = {{Tr{\"o}ster}, Tilman and {Ferguson}, Cameron and {Harnois-D{\'e}raps}, Joachim and {McCarthy}, Ian G.},
        title = "{Painting with baryons: augmenting N-body simulations with gas using deep generative models}",
      journal = {\mnras},
         year = 2019,
        month = jul,
       volume = {487},
       number = {1},
        pages = {L24-L29},
          doi = {10.1093/mnrasl/slz075},
archivePrefix = {arXiv},
       eprint = {1903.12173},
 primaryClass = {astro-ph.CO},
       adsurl = {https://ui.adsabs.harvard.edu/abs/2019MNRAS.487L..24T}
}

\end{document}